# Unraveling Size Dependent Bi- and Tri-exciton Characteristics in CdSe/CdS Core/Shell Quantum Dots via Ensemble Time Gated Heralded Spectroscopy

*Einav Scharf[1], Rotem Liran[1], Adar Levi[1], Omer Alon[1], Nadav Chefetz[1], Dan Oron[2*], Uri Banin[1*]*

[1] Institute of Chemistry and the Center for Nanoscience and Nanotechnology, The Hebrew University of Jerusalem, Jerusalem 91904, Israel

[2] Department of Molecular Chemistry and Materials Science, Weizmann Institute of Science, Rehovot 7610001, Israel

## ABSTRACT

Multiexcitons (MXs) in quantum dots (QDs) manifest many body interactions under quantum confinement. Beyond this fundamental interest, MXs are of importance in numerous optoelectronic applications including QD lasing, light emitting diodes and photocatalysis. Yet, the strong interactions between MXs leading to rapid non-radiative decay introduce challenges for their characterization. While so far, the measurement techniques rely either on indirect methods or on single particle studies, herein we introduce a new method to study MXs in QD ensembles utilizing spectrally resolved time-gated heralded spectroscopy. With this approach we extract the biexciton binding energies in a series of CdSe/CdS QD ensembles of several core/shell sizes, manifesting a transition between attractive and repulsive exciton-exciton interactions. Additionally, for triexcitons, which involve occupation of two excitons in the $1s$ energy levels, as well as one exciton in the $1p$ energy levels, we address the open issues of isolating the spectra of the two triexciton pathways from one another and from high-order MXs, and extract the MX lifetimes. The measurements on ensembles provide high photon counts and low noise levels, and alongside the time-gated heralded approach thus enable the observation of MX characteristics that are difficult to resolve in single particle studies. The approach can be further implemented in the characterization of the energies and lifetimes of MXs in other QD systems to enable rapid characterization and understanding of the MX properties. Such insight bears relevance to optoelectronic applications ranging from lasing to electroluminescent devices to quantum light sources.

## MAIN TEXT

## Introduction

Multiexciton (MX) states in colloidal quantum dots (QDs) serve as an ultimate model system for many body interactions of excitons under strong quantum confined conditions. Unlike molecules, and more like bulk semiconductors, the QD states can occupy numerous excitons in close proximity and with strong overlap, without significantly altering the level structure.[1,2] However, unlike the bulk semiconductor limit, the quantum confinement condition induces strong Coulomb and Exchange interactions with consequences of shifting the energy of MX states relative to the single exciton (X) state, and also invoking efficient Auger-type non-radiative interactions, the rate of which typically scales with the QD volume.[3–5] In strongly confined QDs, non-radiative Auger decay thus often effectively competes and overtakes the radiative decay of the MX states.[6] This imposes limitations on the study and analysis of the energetic and dynamic characteristics of the MX states.

Beyond the fundamental many body physics, understanding the energetics and dynamics bears also important consequences for the utilization of MX states in diverse applications of QDs. First, in lasing, optical gain requires excitation of biexcitons (BXs).[7–10] The hastened Auger decay processes of the BX state though, counteracts the radiative emission essential for lasing and hence numerous avenues, including rod architecture,[11–14] graded shell compositions,[15–19] modifying the core/shell band alignment from type I (straddling) to type II (staggered),[20,21] giant shell QDs ,[22–24] quantum shell systems,[25,26] and most recently also coupled QD molecules,[27,28] have been pursued to manipulate and slow down the Auger decay rates. On the other hand, the increased Auger decay benefits the single photon emission characteristics of single colloidal QDs by quenching the BX emission and leading to strong photon antibunching without any optical filtering.[29] MX states in QDs may also be relevant for light harvesting applications in photovoltaics and photocatalysis.[30,31]

With these motivations in mind, we decipher herein open issues in the energetics and dynamics of MX states in QDs via introduction of a novel spectroscopic approach of ensemble time gated heralded spectroscopy. Considering first the simplest MX case of the BX state, the BX binding energy depends on the net Coulomb interaction between the charge carriers.[32] an attractive (repulsive) X-X interaction leads to a red (blue) shifted BX spectrum and a positive (negative) BX binding energy.[33] In type-I core/shell QDs, the electrons and holes are confined to the core, and the net X-X interaction is usually attractive.[2,33,34] This is also the case for core only QDs, in which the electrons and holes' wave functions highly overlap.[1,10,35] On the other hand, in a type-II band alignment, the reduced attractive interaction between the separated electrons and holes is usually insufficient to overcome the strong repulsion between the same charge carriers (holes/electrons) that occupy the same region, resulting in a repulsive X-X interaction.[21,36,37] Herein we thus investigate CdSe/CdS core/shell QDs as an outstanding model system manifesting the transition between type I to (quasi) type II behavior based on core/shell dimensions.

Observing the X-BX spectral shifts on the ensemble level is challenging as they are on the order of or smaller than the inhomogeneous spectral width, resulting in a significant overlap between the X and BX spectra.[38] Prior studies rely on time-resolved or quasi-continuous-wave (CW) power-dependent photoluminescence and transient absorption measurements.[35,39] Examining the spectrum upon increasing excitation powers allows to follow the evolution of spectral peaks and features,[1,40] as in higher excitation power, the number of Xs increases leading to sequential state filling. On the ensemble level this X occupation will follow the Poisson distribution.[41] Using transient absorption allows to study the spectrum and its dynamics at very short timescales, in which the MX signal is dominant.[42,43] Other works rely on low temperature measurements, which eliminate the thermal broadening, improving the spectral separation between the emitting states, although the inhomogeneous broadening still masks significant information, and further occurrence of charged exciton transitions also complicates the interpretation of the spectra.[44,45]

The case of triexcitons (TXs) is even more intriguing, as a TX state involves occupation of an electron in the $1p_e$ energy level and a hole in the $1p_{3/2}$ level, due to the twofold degeneracy of the $1s_e$ energy level and driven by the repulsion between three holes in the $1s$ states (Figure 1e).[46] The $1p_{3/2} - 1p_e$ optical transition is higher in energy than the band gap X $1s_{3/2} - 1s_e$ transition (marked by the blue and red arrows in Figure 1e, respectively), and is thus well separated from the X spectrum.[10] Accordingly, emergence of a blue shifted peak indicates TX emission.[46] Importantly, the TX emission can also arise from the recombination of an electron in the $1s_e$ energy level (marked by the yellow arrow in Figure 1e), followed by relaxation of the $1p$ electron and hole to the band edge.[47,48] The two TX transitions thus lead to emission in two colors, which can be utilized for two-colored QD lasers.[49,50] Yet, when studying the TX emission, the $1s_{3/2} - 1s_e$ transition is difficult to detect as it nearly overlaps the X and BX spectra.[48]

In order to more directly study MXs, so far usually single particle studies were deemed essential. When exciting single QDs with a pulsed laser, detection of two photons following a single excitation pulse clearly indicates BX emission, while detection of three photons indicates TX emission.[33] Although the arrival time of multiple photons can be detected by splitting the emission signal to multiple single photon avalanche diodes (SPADs), it is challenging to precisely characterize the energies of the emitting states in the standard Hanbury-Brown-Twiss setup or its extensions.[47,51] An estimation of the emission energies can be done by filtering the signal in each SPAD or by diffracting the emission signal and scanning it with a SPAD.[48,52–54] This concept was used to measure the BX binding energy in single CdSe/CdS/ZnS QDs.[52] Addressing the challenging TX state, in another work, the emission signal was split to four spectrally filtered SPADs, which allowed to distinguish between TXs that are emitted from the $1s$ level and TXs that are emitted from the $1p$ level.[48] This study suggested that the $1s_{3/2} - 1s_e$ transition was dominant in the TX emission. This improved the understanding of the energetics of multi-excited states, yet it did not allow to construct the entire MX spectrum.

Lately, the promising method of heralded spectroscopy was introduced to study the electronic characteristics of single quantum dots, significantly improving the characterization of MXs.[33] In this method, the spectrally resolved emitted light is detected by a SPAD array, providing simultaneous spectral and temporal information for each of the emitted photons along with photon statistics. This allows to post-select events where multiple photons arrive in a single excitation pulse and therefore to study the spectrum and lifetime of each emitting state individually. Applying this approach on various single QDs revealed an attractive X-X interaction in single CdSe/CdS/ZnS QDs,[33] resolved the controversy regarding the BX binding energy of CsPbBr$_3$ nanocrystals demonstrating that it is attractive,[39,40,43,55] and found two BX types in coupled QD molecules.[56] Recently, this method was even extended to study TX emission in giant perovskite nanocrystals.[57] These works therefore established the ability of heralded spectroscopy to resolve the ambiguity of previous methods in determining the emission energies of MXs.

These prior studies focused on single QDs to assign the X, BX, and TX in the cascaded photon emission. However, single particle studies require long integration times, ultra-stable QDs, an elaborate statistical study, and the data can be noisy, especially when studying MXs that emit in low quantum yields.[48] Furthermore, they may suffer from selection bias, as bright and stable QDs may not represent well the entire QD ensemble. Recently, a theoretical work laid the groundwork for measuring the TX lifetime and quantum yield in solution, demonstrating an opportunity to extract MX properties from ensemble of QDs.[58] Still, a complete spectral characterization of MXs in ensemble has yet to be addressed.

Herein, we introduce and utilize the ensemble time gated heralded spectroscopy method on small assemblies of QDs, using CdSe/CdS QDs as a model system. This approach is unintuitive as the detected multiple photon events can also originate from uncorrelated emission of Xs from multiple QDs.[58] Yet, we combine the heralded spectroscopy approach with time gating, to disentangle this complexity and obtain signals that are highly dominated by MX emission and selective to specific transitions. Advantageously, with this method, $10^4$-$10^6$ events of photon pairs and photon triplets may be collected within tens of seconds. These high signals improve significantly the photon statistics and signal to noise ratio and enable to extract the BX binding energy, to distinguish between the $1s$ and $1p$ TX emission, and even to observe MXs of higher orders. Additionally, by applying time gating with sub-100 picoseconds (ps) resolution we resolve and find the lifetimes of the MX states, including the lifetimes of the TX decay pathways, a task which has hitherto eluded emission-based studies. Applying this approach on small ensembles of CdSe/CdS QDs of different sizes, we show the crossover from a repulsive X-X interaction in the quasi-type-II regime to an attractive interaction in the type-I regime upon probing large cores with thin shells. Beyond the extraction of the MX characteristics for the model CdSe/CdS QD system, the approach opens up opportunities to study MX states in other QD systems which may be more challenging to address using alternative approaches or the single particle spectroscopy due to lower MX emission quantum yields. Especially considering the challenges of designing optimal hetero-structured QD systems for demanding lasing applications, electroluminescent devices and additional applications, multiple parameters are varied in the structures which requires rapid analysis of their MX characteristics to derive reliable structure-function correlations fed back to QD design.

## Results and discussions

### Time gated heralded multiexciton spectroscopy approach for QD ensembles

A series of CdSe/CdS QDs of various core/shell dimensions (2.2/1.5, 1.4/2.3, 1.8/5.2, 2.2/5.5 nm core radii/shell thickness) were synthesized following a well-established procedure.[59] Figure 1a-d present the transmission electron microscope (TEM) images of the various samples (the same images and the size statistics are in Figure S1 in the supporting information). To measure the optical properties of MXs in the QDs, a solution of CdSe/CdS QDs in 2.5% poly(methyl methacrylate) in toluene, was spin coated on a glass coverslip and the experiments were conducted on an inverted microscope with an oil-immersion objective, as schematically presented in Figure 1f. A widefield photoluminescence (PL) image of the QDs, exemplified for the sample with core/shell radii of 1.8/5.2 nm, respectively, revealed bright spots of aggregates of QDs, along with dim spots of single QDs (Figure S2). To utilize heralded spectroscopy on ensembles of QDs, we focused our study on the QD aggregates. The QD aggregates were placed at the focus of the objective, where they were excited by a pulsed 405 nm laser. The emitted photons were collected with the same objective.

As illustrated in Figure 1f, the emission light is then focused into a spectrograph with a grating that diffracts the emission signal onto a SPAD array detector, such that simultaneous recording of each single photon is recorded alongside its energy. To verify the characteristics of the studied aggregates, a full optical-structural correlation was performed utilizing our previously reported on-chip correlation method (Figure 1g).[60,61] Briefly, an optical characterization on a glass substrate with lithographically predefined markers is followed by extraction of the region of interest with a dual focused ion beam - scanning electron microscope. The extracted area is then processed into a lamella for scanning transmission electron microscope (STEM) characterization. The left image in Figure 1g is an optical widefield image of a 1.8/5.2 nm QD aggregate on a marked substrate. The right image is the bright-field STEM image of the very same aggregate, revealing that it consists of 148 QDs (the complete optical-spectroscopic characterization of this aggregate is presented in Figure S3).

To study MX states on such an ensemble, we measured the PL spectral-temporal characteristics of the aggregates for a series of increased laser excitation powers. Figure 1h presents the power dependent spectra of another 1.8/5.2 nm aggregate. Increasing the excitation power influences the emission spectrum in two manners; 1. The main emission peak blue shifts from 1.95 eV to 1.97 eV, and 2. Additional spectral features evolve in the region above 2.1 eV.

These observations indicate that additional emitting states contribute to the spectrum in higher fluences.

In order to resolve and assign the additional peaks that arise in the spectrum we utilize the two unique abilities of the SPAD array detector: time gated spectroscopy and heralded spectroscopy. Time gated spectroscopy correlates the emission energy with the arrival time of each photon to reveal the fluorescence decay dynamics of different spectral features. Figure 1i, j present two-dimensional spectrum-lifetime plots of the aggregate in Figure 1h under low and high excitation powers (average number of generated Xs per QD per pulse of $\langle N \rangle = 0.1, 6.6$, respectively; the black gaps are due to excluded noisy pixels). At low fluence, a spectral peak near the band gap energy appears, with essentially a long lifetime (Figure 1i). At high fluence (Figure 1j), there are clearly two spectral features: The one near the band gap manifests now much faster decay, while the emergent shoulder peak in high energy exhibits even shorter lifetime. These changes are all due to the emergence of MX emission.

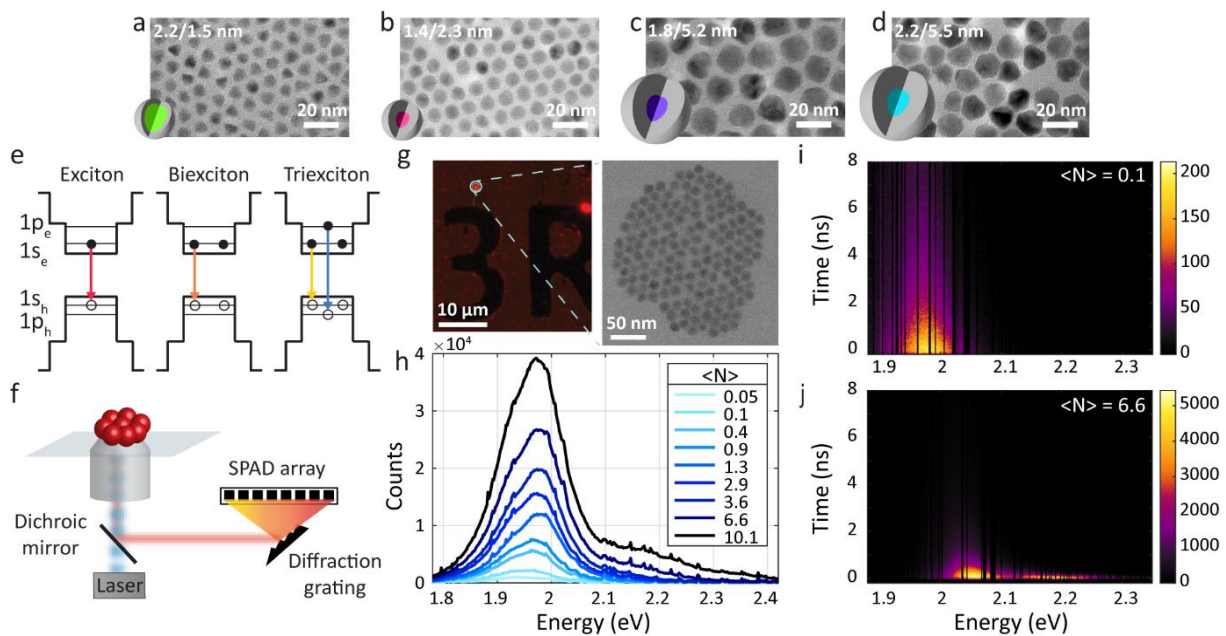

*Figure 1. (a)-(d) TEM images of the studied CdSe/CdS samples. The core/shell radii are 2.2/1.5, 1.4/2.3, 1.8/5.2, and 2.2/5.5 nm. The core/shell illustrations demonstrate the different sizes. (e) Schematic of the configuration of the first energy levels in the QD, showcasing the optical transitions of an exciton (red arrow), biexciton (orange arrow), and triexciton (yellow and blue arrows) states. Notably, in the triexciton state, the $1p$ energy level is occupied and there are two possible transitions: $1s_{3/2} - 1s_e$ in lower energy (yellow arrow) and $1p_{3/2} -$*

*$1p_e$ in higher energy (blue arrow). (f) Illustration of the experimental setup. An aggregate of QDs on a glass coverslip is illuminated by a pulsed laser, focused by an objective. The emitted light is collected by the objective and detected by a SPAD array at the output of the spectrograph. (g) A fluorescence widefield image of the sample in the left panel, showcasing a typical 1.8/5.2 nm QD aggregate, marked by a circle. The right panel displays a bright-field scanning transmission electron microscope (STEM) image of the same aggregate, containing 148 QDs. (h) Power dependent spectra of 1.8/5.2 nm QD ensemble. The ⟨N⟩ values are calculated using the saturation power, extracted from Figure 2e. (i) and (j) Two-dimensional spectrum-lifetime histograms of the ensemble in (h) extracted from the SPAD measurements in low excitation power (i) and high excitation power (j). The black gaps are due to excluded noisy pixels. Brighter colors represent higher photon counts.*

To gain the detailed spectral-temporal information on the MX states, we now implement the powerful heralded spectroscopy approach. To this end, we post-selected *photon pair* events to study BX emission, and *photon triplet* events to study TX emission. Figure 2a presents the spectra of the first and second photons in the post-selected photon pairs (in blue and pink, respectively) in low excitation power, showcasing a blue shifted spectrum of the first photon, which is the BX photon. In previous works on single QDs, this shift was attributed to the BX binding energy, resulting from the repulsive X-X interaction.[52,56] However, in ensemble of QDs this assignment needs to be revisited and examined, as photon pair events may include BXs as well as two uncorrelated X emissions from different QDs. However, in this latter case, both the first and second photons in such pairs would reflect the ensemble inhomogeneous distribution (e.g. different emission due to different QD size), and should not manifest noticeable shifts.

Further insight is obtained by utilizing the time gating capability of the data from the SPAD array. Due to non-radiative Auger processes, as well as the enhanced radiative rate in BX relative to X states, the BX ought to manifest a short lifetime.[62] Thus, to enhance the relative contribution of BX emission in the signal, only photon pairs where the first photon arrived in the first 1 ns after the laser pulse, were post-selected. These temporal conditions are illustrated in Figure 2b, which presents the first 20 ns of the total lifetime measured for this sample (the repetition rate was 2 MHz). The blue area highlights the first 1 ns, as the temporal gate of the arrival time of the first photon, and the pink area presents the arrival time of the second photon, which is limited to 200 ns after the first photon to reduce contribution of dark counts (the full temporal conditions are detailed in the Methods section). Notably, the BX energies, obtained from this analysis, are robust to small changes in these temporal gates. For example, increasing the chosen temporal gate of the first photons by 50%, changes the peak energy by less than 1 meV (Figure S4).

A similar strategy allows us to extract spectroscopic information also for TX states, which are much more difficult to study on the single particle level due to the low number of such emissive events in typical QD systems. To this end, Figure 2c presents the spectra of post-selected photon triplets in higher excitation power. Here the first photon, in purple, represents the TX emission (temporal gate of 1 ns; purple area in Figure 2d). The second and third photons represent the BX and X emission in blue and pink, respectively (temporal gates of 3, 200 ns, in the blue and pink areas in Figure 2d, respectively). The BX spectrum is still blue shifted from the X spectrum, as in the photon pairs spectra at low power, validating further this analysis. The X emission (in pink) has a more complex spectrum shape, as it consists of multiple emitting states, such as charged Xs and also BXs. Interestingly, the TX emission features two distinct peaks, both are blue shifted from the X and BX peaks. The high energy peak, which is associated to emission from the $1p$ state, is visibly enhanced in comparison to the full spectrum in gray, clearly showcasing the significance of the post-selection used in heralded spectroscopy for the observation of MXs.

Figure 2e presents the number of single photons, photon pairs, and photon triplets per second (in the top, middle, and bottom panels, respectively), as a function of the excitation power. In the three highest excitation powers the counts start to rise, assigned to transition to a quasi-CW regime, where the multi-excited QDs emit rapidly during the laser pulse (Figure S5),[23] hence they are neglected in this analysis. The power dependence plots are fitted to the probabilities to generate at least one, two, or three Xs, respectively, according to the Poisson distribution.[10] The saturation power is extracted as a fitting parameter from the single photons' fit in the top panel and is later used to fit the photon pairs and triplets in the middle and bottom panels. This shows that by applying a short time gating on the arrival times of the first photons, the detection ratio of MXs to multiple Xs is sufficiently high to produce a good fit.

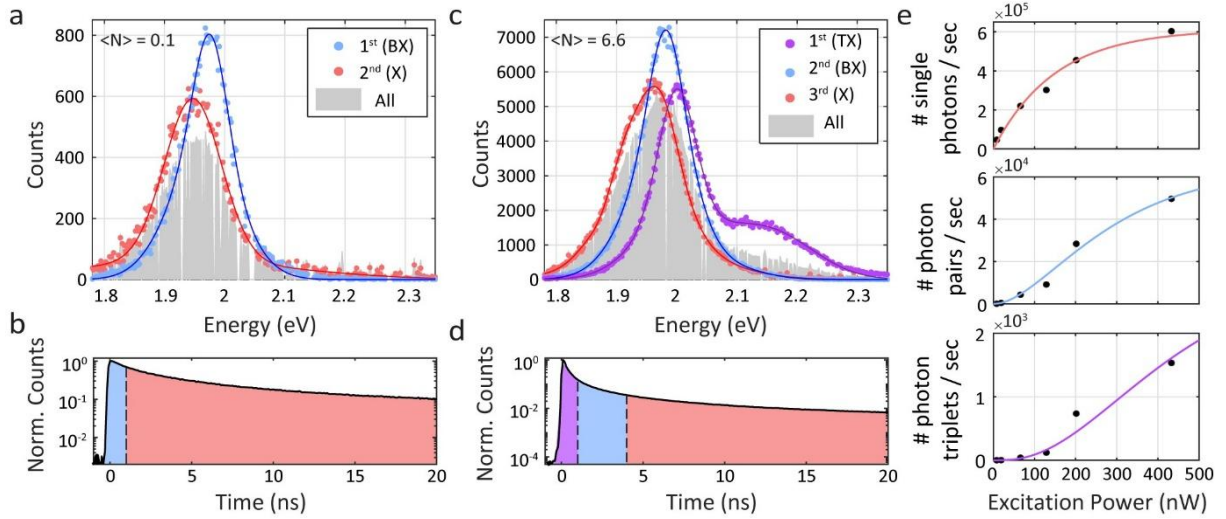

*Figure 2.* Time gated heralded spectroscopy in QD ensembles. (a) Spectral characterization of the photon pair events under low excitation power. The blue dots represent the first photons in the pairs ('BX') and the pink dots represent the second photons in the pairs ('X'). They are both fitted to a multi-gaussian model (in blue and red, respectively). The gray area is a normalized spectrum of all photon events. The gaps are due to excluded noisy pixels. (b) The first 20 ns of the total lifetime in low power ($\langle N \rangle$=0.1). The colored areas mark the temporal conditions applied on the arrival time of the photon pairs. Accordingly, the photon pair spectra in (a) only feature pairs where the first photon arrived in the first 1 ns after the laser pulse (as marked by a blue area), and the second photon follows (marked by a pink area). (c) Spectral characterization of the photon triplet events under high excitation power. The first (TX, 1 ns arrival gate), second (BX, 3 ns gate), and third photons (X, 200 ns arrival gate) are in purple, blue, and pink dots, respectively, and are fitted to a multi-gaussian model. The gray area is a normalized spectrum of all photon events. The temporal gates for the arrival times of the first and second photons are shown by the purple and blue areas under the total lifetime in this excitation power in (d). (e) The number of single photon events, photon pair events, and photon triplet events per second versus excitation power in the top, middle, and bottom panels, respectively (in black dots). The colored lines are fits to the probability to generate at least one, two, or three Xs (in pink, blue, and purple, respectively) according to the Poisson distribution. The saturation power is extracted from the fit of the top panel and used in the fits of the middle and bottom panels. Then $\langle N \rangle$ is the ratio between the used power and the saturation power.

In order to further analyze and discuss the BX behavior in the QD ensemble, we first characterize fully the emission spectrum under low excitation fluence ($\langle N \rangle = 0.1$), where the probability to generate TXs is significantly lower than the probability to generate BXs (by more than an order of magnitude). To extract the BX binding energy, we performed a time gated heralded analysis on the photon pair events, such that the first photons arrive in 2 ns bins and the second photons arrive later. Figure S6 presents the emission peaks of the first and second photons as a function of time, interestingly, both red shifting with time. Figure 3a-b demonstrate the spectra, achieved by this analysis, exemplified for the case where the first photons arrive in the first 2 ns (Figure 3b) and the second photons follow (Figure 3a). These spectra are well fitted to a multi-gaussian model. To understand the contributing emitting states, we followed the emission of a single QD in low excitation power and found that its emission peak redshifts with time as well (Figure S7, 8). By following the energies of the "on" and "off" states within its emission intensity trace (likely representing neutral and charged X emission),[63] we were able to attribute slower (faster) and red (blue) shifted emission to the neutral (charged) X emission. Therefore, the apparent red shift with time in the single QD is a result of the transition between these two emitting states. Hence, the charged X state must be considered in the analysis of the emitting states in the QD ensemble as well.

Consequently, we fitted the spectrum of the second arriving photons in the pairs (corresponding to the singly excited state) to two gaussian components (Figure 3a), one in lower energy (1.942 eV, in black), to represent the neutral X, according to the energy of the X in longer times (Figure S9), and another peak in higher energy (1.960 eV, in red) to represent the charged X emission. The first photons' spectrum was fitted to three gaussian components (Figure 3b), one in high energy (1.979 eV, in blue) according to the energy of the first photons in short arrival times (Figure S4), to represent the BX emission, and the two X and charged X components from Figure 3a, to account for emission from two singly excited QDs. This multi-gaussian fit was applied for each of the time-dependent spectra of the second and first photons, in order to follow the temporal evolution of the emitting states.

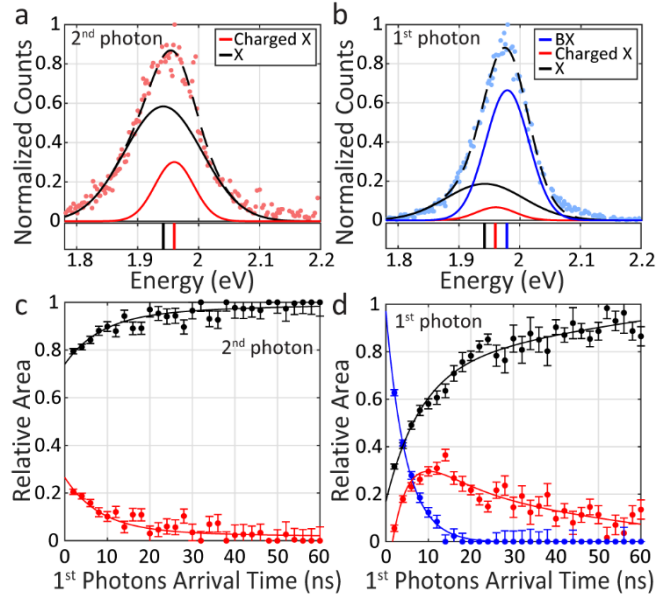

***Figure 3.*** *Biexciton characterization in QD ensembles. (a), (b) The spectrum of the second and first photons (in pink and blue dots, respectively) within the photon pair events at low excitation power of $\langle N \rangle = 0.1$. Only pairs where the first photon arrived in the first 2 ns are included. The second photons in (a) are fitted to a sum of two gaussians (in a dashed black line). Its components represent neutral X emission (black line) and charged X emission (red line), and their peak energies are marked with the same color code under the spectra. Similarly, the first photons' spectrum in (b) is fitted to a sum of three gaussians (in a dashed black line). The black and red components are the same as in (a), representing neutral and charged X emission, respectively. The blue line represents BX emission. (c) and (d) present the temporal evolution of these states (bins of 2 ns for the arrival time of the first photons in the pairs), as the relative integrated area under each fitted gaussian in the spectra of the second and first photons (in c and d, respectively). In the second photons in c the contribution of the X increases and that of the charged X decreases. In the first photons in d the contributions of the X and charged X increase and that of the BX decreases rapidly. Later the contribution of the charged X slightly drops. The lines are multi-exponential fits.*

Figure 3c presents the relative integrated area under the fitted gaussian components of the neutral X (in black) and the charged X (in red) versus time for the second photons in the pairs (following the example in Figure 3a). It shows a simultaneous decrease in the relative contribution of the charged X and growth in the relative contribution of the neutral X, in accordance with the longer lifetime of the neutral X. To extract the rates of the changing contributions of the neutral and charged Xs, we applied a bi-exponential fit, following a first-order kinetic model. We used this analysis to estimate the lifetime of the charged X state.

Therefore, the rate constant of the slow component was fixed to 1/105 ns$^{-1}$, as 105 ns is the lifetime of the neutral X according to the total lifetime of all photons at $\langle N \rangle = 0.1$, presented in Figure S10 (the total lifetime in Figure S10 exhibits three lifetime components of 2.4±0.1, 13.8±0.2, 105±1 ns). The bi-exponential fit of the time-dependent relative area contributions, yielded a rate constant, analogous to a lifetime of 8±1 ns for the fast component. This rate is similar to the intermediate lifetime component in Figure S10, which was previously attributed to charged X emission.[28]

Figure 3d presents the relative integrated area under each of the gaussian components within the first photons' emission (following the example in Figure 3b). It exhibits a fast decay of the BXs' relative contribution (in blue), along with an increase in the relative contribution of the other two peaks. Then there is a slight decrease in the contribution of the charged X (in red). A mono-exponential fit to the BX curve yielded a rate constant, corresponding to a lifetime of 4.7±0.1 ns, which is similar to the short lifetime component of all photons (Figure S10), attributed mostly to BXs at the weak excitation regime.[28] The X curve was fitted to a tri-exponential fit with rate constants of 1/4.7, 1/8, 1/105 ns$^{-1}$, reflecting the influence of all the emitting states on the change in the relative contribution of the X component. The dynamics of the relative contributions of the states in Figure 3d resemble the kinetics of consecutive first order reactions. Therefore, we treated the charged X state as the intermediate in such mechanism. Its curve was fitted to a difference between exponents, with formation component with a rate of 1/4.7 ns$^{-1}$, according to the BX rate, and decaying components with rates of 1/8, 1/105 ns$^{-1}$, according to the neutral and charged X rates. Interestingly, the same initial rate for the decay of the BX and the formation of the X components showcases that the BX state relaxes either into the neutral or the charged X states. Eventually, in longer times, the charged X component decays as well, as the neutral X component becomes the dominant emitting state.

**Extracting quantitative binding energies and decay kinetics – the case of biexcitons**

Following this understanding of the BX behavior in the aggregates, we extract the BX binding energy, the energy difference between the neutral X emission and the BX emission. We obtain a repulsive BX binding energy of -37±4 meV for this ensemble.

We next utilize time-gated heralded spectroscopy in QD ensembles to study and to accurately determine the BX binding energy, for a series of CdSe/CdS QDs of various sizes (TEM images in Figure 1 and Figure S1). Given the freedom in tuning either core or shell sizes, it is clear that the rapid study enabled by the ensemble gated heralded spectroscopy is essential here. Figure 4a-d demonstrate the spectra of the first and second photons within the post-selected photon pairs for the various samples. The sizes are noted in the top left corner of each panel, as the core/shell radii. Notably, the aggregates in Figure 4b-d showcase a blue shifted emission of the first photons' (BX) spectrum, whereas Figure 4a showcases a red shifted emission of the first photons' spectrum.

Figure 4e presents a summary of the BX binding energies of all the QD ensembles as a function of the X energy. For the larger 2.2/5.5, 1.8/5.2 nm QDs the BX binding energy was extracted following the analysis introduced above (Figure 3). For the smaller 2.2/1.5, 1.4/2.3 nm QDs the X energy shift with time was smaller, thus two components of neutral X and BX were sufficient to fit the data and extract the BX binding energies (Figure S11 presents example of this analysis, Figure S12 summarizes the charged X energies of the two larger QD samples). Additionally, a higher excitation power was required to observe sufficient BX signal in the smaller QDs. Thus $\langle N \rangle \sim 1$ was used (instead of $\langle N \rangle \sim 0.1$ for the larger QDs). The average BX binding energies were 2±1, -13±3, -29±3, -37±5 meV for the QDs of sizes 2.2/1.5, 1.4/2.3, 2.2/5.5, 1.8/5.2 nm, respectively. In prior work we studied 1.35/2.1 nm QDs on a single particle level and found a BX binding energy of -14±6 meV, in line with the BX binding energy of the similar sized 1.4/2.3 nm QD ensemble study performed herein.[56] Additionally, measurements of single 1.8/5.2 nm QDs revealed a BX binding energy of -31±9 meV (Figure S13). The similarity between the results of single QDs and ensembles of these sizes of QDs further supports the validity of utilizing heralded spectroscopy in small QD ensembles.

The size dependence of the BX binding energy is attributed to the quasi-type-II band alignment of the CdSe/CdS QDs (illustration of the charge carriers' wave functions in a quasi-type-II system is in the inset of Figure 4f). On the one hand, when the core is bigger, the confinement of the holes decreases, reducing the X-X repulsion. On the other hand, when the shell is thicker, the much lighter electrons experiencing also a very small core/shell conduction band-offset potential, delocalize further and the attractive interaction between the electrons and holes is reduced, enhancing the apparent X-X repulsion. Accordingly, in the transition from 1.4/2.3 nm QDs to 1.8/5.2 nm QDs the interaction becomes more repulsive, showcasing dominance of the effect of the thicker shell. To elucidate this dual size dependence, the BX binding energies can be compared to the ratio of the core radius to shell thickness, showcasing a consistent reduction in the repulsion with the increasing ratio (Figure 4f).

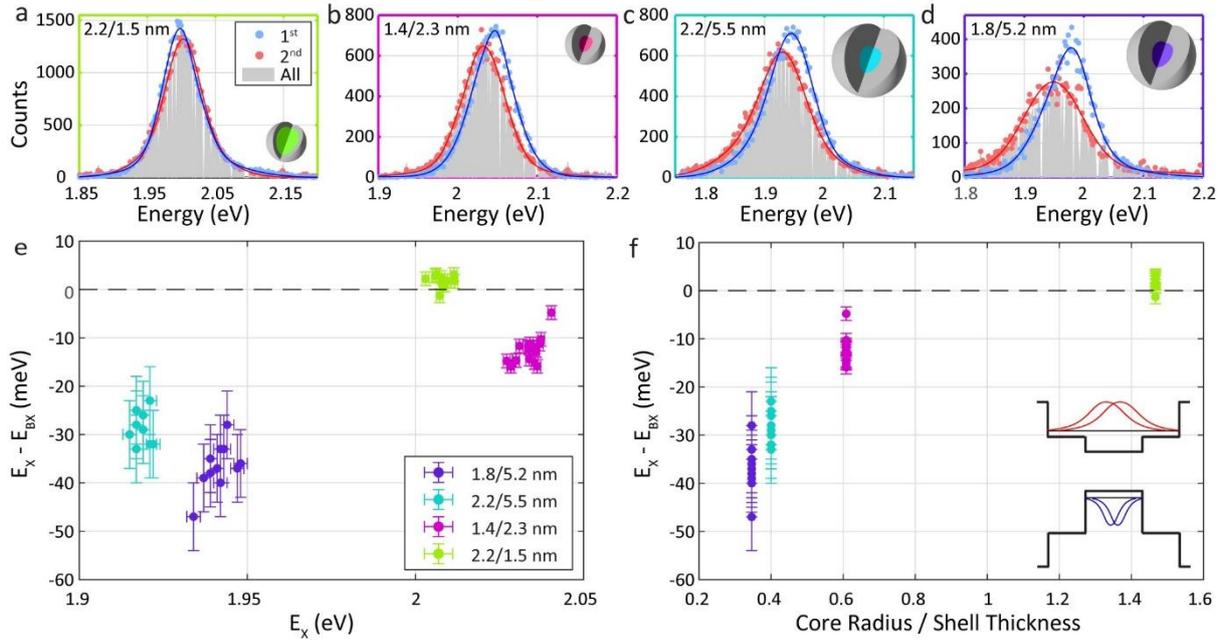

***Figure 4.*** *Size dependent behavior of multiexcitons in CdSe/CdS core/shell QDs. (a) to (d) present the photon pair spectra of samples from different sizes of CdSe/CdS QDs (core/shell illustrations demonstrate the different sizes). The spectrum of the first (BX) photons (in blue) is blue shifted from the spectrum of the second (X) photons (in pink), in the sizes 1.4/2.3, 2.2/5.5, and 1.8/5.2 nm (in b, c, and d, respectively). The 2.2/1.5 nm QDs exhibit a red shifted spectrum of the first photons, reflecting dominant attractive interaction. The gray area is a normalized spectrum of all photon events. The gaps are due to excluded noisy pixels. (e) The BX binding energy of the various QD ensembles versus the X energy. (f) The BX binding energy of the various QD ensembles versus the ratio between the core radius and the shell thickness. The BX binding energy becomes less negative in higher ratio. Inset: illustration of a BX in a quasi-type-II core/shell band alignment. The holes in the valence band are confined to the core (in blue) and the electrons in the conduction band are delocalized to the shell as well (in red).*

As mentioned, QDs of 2.2/1.5 nm exhibited an attractive BX binding energy. Comparing these QDs to the QDs of 2.2/5.5 nm, with the same core size, which exhibit strong X-X repulsion, reveals yet again a significant dependence on the shell thickness. When the shell thickness is reduced from 5.5 to 1.5 nm, the electrons localize closer to the holes in the core, which enhances the attractive interaction between them. The overall attractive X-X energy is consistent with the typical BX binding energy in type I core/shell QDs.[33,34] Measuring an attractive BX binding energy is significant, as it showcases the ability of this method to observe the crossover from a quasi-type-II to a type-I band alignment.[37] Moreover, in an ensemble, detection of two photons where the first one is in higher energy might be confused with other effects, such as the size-dependent quantum confinement and energy transfer.[64,65] Yet, observation of two photons where the first photon is in lower energy cannot be explained by these phenomena, further solidifying the validity of this technique.

**Analysis of triexciton energetics and branching emission dynamics**

Following the quantitative understanding of the BX emission in the QD ensembles, we analyze the behavior of the 1.8/5.2 nm QD ensemble, presented in Figures 1-3, under higher excitation fluence ($\langle N \rangle = 6.6$) in order to understand the TX emission. We applied a similar approach with a multi-gaussian fit to the spectrum, combining heralded spectroscopy analysis with time gating. The spectrum of the first photons within the photon triplets, as appears in purple in Figure 2c, was fitted to 5 peaks (Figure 5a). The red and blue peaks represent the X and BX emission, respectively. As in the BX analysis, the first photons within the triplets may exhibit contribution of emission of Xs and BXs, due to multiple photons detection of emission from different QDs within a single excitation pulse. The single exciton peak energy extracted in this high excitation fluence is 1.960 eV (relative to 1.942 eV at lower fluence) as higher contribution of charging is expected. Indeed, the spectrum of the third photons in the triplets (pink spectrum in Figure 2c) is blue shifted relative to the X spectrum in lower power (pink spectrum in Figure 2a). However, to maintain a reasonable number of peaks we prefer to use in the fit procedure a slightly modified position for this effective 'X' peak, at 1.960 eV.

Focusing on the first (TX) photons within the triplets, besides the X and BX components, the spectrum is well fitted to three additional peaks at higher energies (centered at 2.01 eV in green, 2.1 eV in magenta, and 2.175 eV in cyan). In order to assign these peaks, we performed time gated analysis for the photon triplets. Figure 5c presents the temporal evolution of the emission spectrum of the first photons in the triplets where each spectrum represents a 20 ps bin of the first 1 ns after the pulse, going from t=0 in light purple to t=1 ns in dark purple (the time-gated build-up of the spectrum in bins of 20 ps is in Figure S14). The higher energy peak, associated with high-order MX emission (TX and above) according to its high energy relative to the band edge peak,[10] decays rapidly, likely due to the fast relaxation of these multi-excited states. The lower energy peak red shifts, due to the relaxation of the blue shifted BX photons.

Similarly, Figure 5e (g) presents the time-dependent spectra of the second (third) photons within the photon triplets in the range of t=0.4-3 ns (t=1-3 ns), ranging from light to dark blue (pink), in time bins of 50 ps (100 ps; For further details see section S2 in the supporting information). The two vertical lines in the spectra in the range of 2-2.1 eV are detector artifacts due to excluded noisy pixels. The spectra were fit to the 5 gaussians as in Figure 5a. Figure 5d, f, h present the results of the temporal evolution of the intensity of each peak for the first, second, and third photons within the triplets, respectively (the color code is the same as in panels a and b; the time bins are 20 ns in panels d, f and 50 ps in panel h). The contribution of the low energy component in the first photons' spectrum was negligible, and was thus disregarded in Figure 5d (the same was done for the high energy component in the second photons in Figure 5f and the three higher energy components in the third photons in Figure 5h).

Fitting these curves allows to extract the rate constants, reflecting the lifetimes of each of the spectral features in Figure 5a. Notably, a mono-exponential decay is detected for each peak, further validating the analysis that yields the well-distinguished emitting states (see table S1 in the supporting information for all the extracted lifetimes). The ability to isolate these states is significant when considering the large extent of their spectral overlap, as appears in Figure 5a.

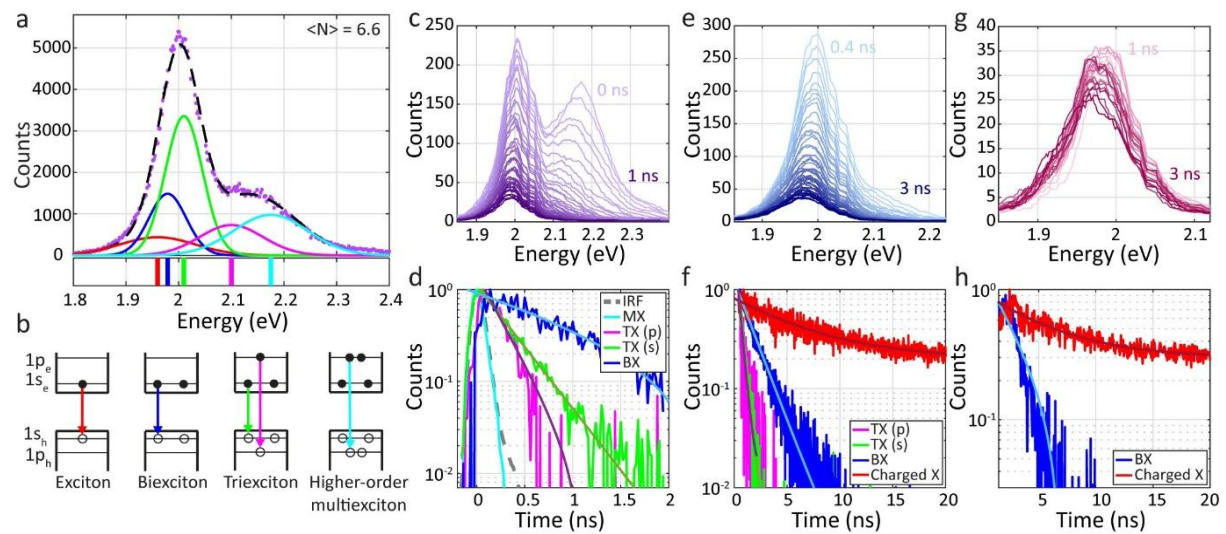

*Figure 5.* *Time gated heralded spectroscopy characterization of multiexcitons. (a) Spectrum of the first photons within the photon triplets (from the analysis in Figure 2c) in purple dots, fitted to a sum of 5 gaussians (in dashed black line: red – X and charged X; blue – BX; green, magenta, and cyan - representing MXs of higher order, TX and more). The peak energies are marked by lines below the spectrum. (b) Illustration with the assignments of the emitting states of each of the 5 gaussians in (a). The red and blue peaks are attributed to X and BX, respectively. The green and magenta peaks are TX emitted from the $1s_{3/2} - 1s_e$ and $1p_{3/2} - 1p_e$ transitions, respectively. The cyan peak is a MX of higher order, for example a tetraexciton,*

*as appears in the illustration. (c), (e), (g) Time gated spectra of the first, second, and third photons within the photon triplets, respectively (the two vertical lines in the range of 2-2.1 eV are detector artifacts due to excluded noisy pixels). The presented temporal ranges for the first, second, and third photons are 0-1 ns, 0.4-3 ns, and 1-3 ns, respectively, going from light to dark colors. The time bins are 20, 50, and 100 ps for the first, second, and third photons, respectively. Each time-dependent spectrum is fitted to the 5 gaussians in (a). (d), (f), (h) Temporal evolution of the emitting states in the first, second, and third photons, respectively (time bins of 20 ps for the first and second photons and 50 ps for the third photons), fitted to mono-exponential decays. The dashed gray line in (d) is the instrument response function. The color code of the decay curves is consistent with the peaks in (a) and the assignment in (b).*

Following these findings we assign the peaks: First, the red peak at 1.960 eV represents the X or the charged X, as illustrated by the red arrow in Figure 5b. Its lifetime of 5.4±0.4 ns can be extracted from the time gated analysis of the third photons within the photon triplets (Figure 5h; or 6.1±0.2 ns according to the second photons in Figure 5f). Fitting the total lifetime of all the photons in this excitation power ($\langle N \rangle$ = 6.6) reveals a triexponential decay with lifetimes of 0.70±0.01, 8±1, 100±45 ns (Figure S10), where the long component has a large error as it represents less than 1% of the decay behavior, as is characteristic of "delayed emission".[66] The lifetime of the X from the time gated analysis is similar to the lifetime of the intermediate component in the total lifetime, supporting its assignment as a charged X.[28] This indicates that in high excitation powers, there should be a contribution of charged MX emission as well, yet it may be less emissive due to non-radiative Auger decay.

The blue peak in Figure 5a corresponds to the BX state, as illustrated by the blue arrow in Figure 5b. Its contribution appears in the first, second, and third photons' spectra (Figure 5d, f, h), and its average lifetime is 2±1 ns (detailed results in table S1 in the supporting information). This lifetime fits the fast component in the total lifetime of all photons at $\langle N \rangle$ = 0.1 (2.4±0.1 ns; Figure S10). This component is dominated by BX emission, as in this power the probability to generate MXs of higher orders is low.

From the decays of the first photons within the triplets (Figure 5d), the lifetimes of the peaks centered at 2.01 and 2.1 eV were 323±5 and 227±7 ps, respectively. The lifetime of the peak centered at 2.175 eV was shorter, similar to the instrument response function (in a dashed gray line in Figure 5d). From the second photons' spectral-temporal analysis (Figure 5f), we extract close lifetimes of 428±3, 360±10 ps for the 2.01 and 2.1 eV transitions, respectively, consistent with contribution of the high energy components also to this signal. The slightly longer values are expected considering the preselection of the time-gated analysis, taking into account only photons with arrival time above 0.4 ns.

Finding such similar lifetimes for the 2.01 and 2.1 eV components is striking, especially without applying any constraints. It showcases that these components reflect indeed a decay from the same initial triexciton state. Therefore, we attribute the 2.01 eV component to emission of a TX through the $1s_{3/2} - 1s_e$ transition and the 2.1 eV component to emission of a TX through the $1p_{3/2} - 1p_e$ transition (green and magenta arrows in Figure 5b). These states should exhibit a similar observed lifetime as upon decay of the TX, either from the $1s$ or the $1p$ state, the remaining BX state only occupies the $1s$ level, due to immediate (~1 ps) relaxation of the $1p$ exciton to the band edge.[67] This is indeed well reflected in the close values extracted as the decay rates of both triexciton components.

The stark difference between the shorter lifetime of the 2.175 eV component and the similar longer lifetimes of the 2.01 and 2.1 eV TX components showcases that the high energy component (2.175 eV) reflects emission of higher-order of MXs. The cyan arrow in Figure 5b demonstrates this assignment as a tetraexciton state, yet this state involves multiple high-order MX states that cannot be temporally resolved by this method, as their fast lifetimes exceed the resolution of the detector. Nonetheless, we were able to distinguish between emission of TXs and MXs of higher-order, which is very difficult in emission-based measurements.

Examining the energy separation between the X and the two TX components provides further validation to their assignments (the energy spacings between all the emitting states in the 1.8/5.2 and 2.2/5.5 QDs are in Figure S15). First, the separation of the $1p$ peak (at 2.1 eV) from the $1s$ X peak (at 1.942 eV) is in line with the literature.[48,49,68,69] Second, the $1s$ TX emission is blue shifted by 68 meV from the neutral X peak at 1.942 eV, which is almost twice higher than the blue shift of the BX emission (37 meV). This shows that an additional X occupation leads to a similar effect of X-X interactions. Yet, the additional repulsion, in a TX state relative to the BX state, is slightly weaker (additional 31 meV) due to lower overlap between the wave functions of the charge carriers in the $1p$ state and charge carrier in the $1s$ state.

The power dependent study of the QD ensemble allowed to follow the evolution of the MX states with the excitation power. Interestingly, plotting the integrated area under each of the two TX gaussians as a function of the excitation power fits well to the probability to generate at least three Xs, according to the Poisson distribution (Figure S16a). Additionally, plotting the area versus power on a logarithmic scale reveals linear trends with similar slopes (2.0±0.2 and 1.9±0.2 for the 2.01 and 2.1 eV peaks, respectively, see Figure S16b), where the quadratic growth with excitation power, rather than cubic growth, is due to the onset of emission of MXs of higher order.[45] In contrast, the onset of the peak at 2.175 eV is in higher excitation power, it does not fit well to the probability to generate at least three Xs, and the log-log behavior of the area versus power showcases a larger slope (3.5±0.7, Figure S16). This yet again solidifies the assignment of the 2.01, 2.1 eV components as TX and the 2.175 eV component as a MX of higher order.

From this analysis, we find that the $1s$ TX emission is stronger than the $1p$ TX transition, with a contribution of 75%-25%, respectively. This contribution is independent of the excitation power (Figure S17), in line with branching decay kinetics with two competing emission pathways from the same electronic TX state. The dominance of the $1s$ TX pathway stems from its higher rate due to the higher degeneracy in the $1s$ state, with 2-4 recombination pathways, in contrast to the single recombination pathway of the exciton in the $1p$ level. Another factor that affects the ratio of the competing TX emitting pathways is the overlap between the electron and hole wave functions.[48] Simulating the geometry of the 1.8/5.2 nm core/shell QD and solving the Schrödinger-Poisson equations for single excitons using an effective mass approximation (following our previously reported procedure),[70] yielded an electron-hole overlap integral of ~0.6 for an X in the $1s$ state and ~0.2 for an X in the $1p$ state, neglecting multi-carrier interactions. The higher overlap between the charge carriers in the $1s$ state relative to the charge carriers in the $1p$ state, contributes to the dominant $1s_{3/2} - 1s_e$ TX transition. Considering the observed lifetime of the TX states (averaging the two extracted lifetimes yields 275 ps) and the branching $1p$ to $1s$ ratio, we can extract the lifetimes of the two TX transitions (details in section S2 in the supporting information). The lifetimes for the $1s_{3/2} - 1s_e$ and $1p_{3/2} - 1p_e$ TX transitions are 0.37 and 1.1 ns, respectively.

Our results are somewhat different than the previous elegant work on single QDs, which reported that the $1s$ TX emission dominates the emission entirely.[48] In this work, the $1s$ versus $1p$ TX pathways were distinguished by spectrally filtering the MX signal. Accordingly, the photon statistics depend on the spectral position of the band-pass filters and the purity of the emitting states. For example, positioning the filter for the $1p_{3/2} - 1p_e$ transition in energy which is too high can bias the signal to detection of higher order of MXs, which emit in lower quantum yields. This can decrease the apparent contribution of the $1p$ TX state. In contrast, our proposed method obtains the full spectrum of the MX states providing improved accuracy.

Finally, following the assignment of the high energy peak as a high-order MX, the heralded spectroscopy method can be further extended to resolve events of 4 photons emission within a single laser pulse. Figure S18 presents such a case for the 1.8/5.2 nm QD sample. Interestingly, the first photons, exhibit a large contribution of the peak at 2.175 eV, along with a lower energy peak that can contain contribution of the $1s$ transitions of the tetraexciton. The second photons, mostly representing the TX emission, showcase a significant contribution of the peaks at 2.01 and 2.1 eV. Going to higher excitation power of $\langle N \rangle = 10.1$ even allows to distinguish 5 photon events, revealing two additional peaks centered at 2.25, 2.35 eV, representing much higher order MXs (Figure S18). This example further demonstrates the strength of heralded spectroscopy in resolving MX states. Transitioning from single particles to small ensembles provides high photon rates, which enables to observe higher emitting states, and perform time gating in high temporal resolution. Combining this with the careful time-gated analysis, provides reliable spectral-temporal information for each QD sample.

## Conclusions

To conclude, we demonstrated a new approach to utilize heralded spectroscopy for the study of MXs in QDs. By using this technique applied to small QD ensembles, we found the BX binding energies of CdSe/CdS QDs of several sizes, demonstrating a wide range of X-X interaction strengths. Additionally, we characterized and assigned the two competing pathways of TX emission, and also observed MXs of higher order. Usually, these properties were so far tackled utilizing measurements of single particles. In this work, we used the time gating alongside the heralding technique as a "magnifying glass", allowing to observe single particle properties on the ensemble level. Moreover, replacing single particle studies with an ensemble characterization produced much higher photon rates, which diminished the typical noise in single particle measurements. Following this method can effectively yield MX properties of different QD systems with numerous design variables related to the heterostructure architecture, and can be applied even on systems in which the single particle characterization is more challenging. This will allow to obtain reliable data on the spectroscopy and dynamics of MX states in complex QD systems, with multiple variables such as core/shell dimensions etc. Such information can aid to achieve design principles for QDs tailored towards optoelectronic applications where MX characteristics are of relevance, including in lasing, light emitting diodes, single photon and quantum light sources, and even in photocatalysis.

METHODS

**Optical setup:** The measurements were performed with an inverted microscope (Nikon ECLIPSE Ti) in the epi-luminescence configuration. An oil-immersion objective (×100 with a numerical aperture of 1.4, Nikon Achromat) was used to focus the excitation light and collect the emission signal. A 370 nm fibre-coupled light emitting diode (Prizmatix) was used for widefield imaging for the selection of the aggregates. A 405 nm pulsed laser (EPL405, Edinburgh instruments) was used for point illumination (repetition rate of 2 MHz). Back-scattered laser light was filtered by a 425 nm long pass dichroic mirror (DMLP425R, Thorlabs) and a long pass 460 nm filter (ET460lp, Chroma). Then the light was focused to the slit of a spectrograph (Acton SP-2150i, Princeton Instruments) with a telescope (AC254-300-A, AC254-100-AB, Thorlabs). The light was diffracted by a grating in the spectrograph (300 grooves/mm, blazed at 500 nm), then detected by a SPAD array (SPADλ, PI Imaging). The SPAD array was triggered by the laser, using a constant fraction discriminator (FLIM LABS) to turn the laser's NIM output to TTL. The instrument response function was 150 ps (gray dashed line in Figure 5d).

**Time gating:** Time gating was used to enhance the contribution of MXs in the post-selected multiple photon events, as well as to decrease the contribution of dark counts and avoid cross-talk.[71] Accordingly, photon pairs are considered only if the first photon arrived in the first 1 ns (or 2 ns in some analyses) after the laser pulse and the second photon arrived between 0.8 ns to 200 ns after the first photon. The short time gate of the first photon is used to reduce the cases of detection of two Xs, that exhibit a longer lifetime (further details on the time gate of the first photon are in Figure S4). The lower time gate of the second photon is in order to diminish the contribution of cross-talk events and the higher time gate is to reduce the contribution of dark counts. For the smaller 2.2/1.5 and 1.4/2.3 nm QDs, which exhibited a shorter lifetime, a time gate of 60 ns was used for the second photons (instead of 200 ns). Applying the same reasoning, photon triplets are considered if the first photon arrived in the first 1 ns, the second photon followed it in times between 0.8-3 ns, and the third photon arrived in times of 0.8-200 after the second photon.

**Electron microscopy:** The image in Figure 1c is a bright-field STEM image, acquired using a TITAN-THEMIS transmission electron microscope operated at 300 kV. The images in Figure 1a-d are bright-field TEM images, acquired using a Tecnai Spirit G2 operating at an acceleration voltage of 120 kV.

## ASSOCIATED CONTENT

The **Supporting Information** is available free of charge on the ACS Publications website.

Size statistics of the studied quantum dots and supporting experimental analyses.

## AUTHOR INFORMATION

**Corresponding authors**

**Uri Banin -** Institute of Chemistry and The Center for Nanoscience and Nanotechnology, The Hebrew University of Jerusalem, Jerusalem 91904, Israel; E-mail: uri.banin@mail.huji.ac.il

**Dan Oron** - Department of Molecular Chemistry and Materials Science, Weizmann Institute of Science, Rehovot 7610001, Israel; E-mail: dan.oron@weizmann.ac.il

**Authors**

**Einav Scharf** - Institute of Chemistry and the Center for Nanoscience and Nanotechnology, The Hebrew University of Jerusalem, Jerusalem 91904, Israel

**Rotem Liran** - Institute of Chemistry and the Center for Nanoscience and Nanotechnology, The Hebrew University of Jerusalem, Jerusalem 91904, Israel

**Adar Levi** - Institute of Chemistry and the Center for Nanoscience and Nanotechnology, The Hebrew University of Jerusalem, Jerusalem 91904, Israel

**Omer Alon** - Institute of Chemistry and the Center for Nanoscience and Nanotechnology, The Hebrew University of Jerusalem, Jerusalem 91904, Israel

**Nadav Chefetz** - Institute of Chemistry and the Center for Nanoscience and Nanotechnology, The Hebrew University of Jerusalem, Jerusalem 91904, Israel

**Author Contributions**

The manuscript was written through contributions of all authors. All authors have given approval to the final version of the manuscript.

**Funding Sources**

This research was supported by the Israel Science Foundation within the MAPATS program (U. B., grant No. 2655/23).

**Notes**

The authors declare no competing financial interest.

ACKNOWLEDGEMENT

U. B. acknowledges support from the Israel Science Foundation within the MAPATS program (Grant No. 2655/23) and thanks the Alfred & Erica Larisch Memorial Chair. E. S. acknowledges the support from the Hebrew University Center for Nanoscience and Nanotechnology, and from the PBC of the Council of Higher Education. We acknowledge S. Remennik and A. Vakahi for their meaningful contribution in the electron microscopy characterization.

# Unraveling Size Dependent Bi- and Tri-exciton Characteristics in CdSe/CdS Core/Shell Quantum Dots via Ensemble Time Gated Heralded Spectroscopy

Einav Scharf[1], Rotem Liran[1], Adar Levi[1], Omer Alon[1], Nadav Chefetz[1], Dan Oron[2*], Uri Banin[1*]

[1]*Institute of Chemistry and the Center for Nanoscience and Nanotechnology, The Hebrew University of Jerusalem, Jerusalem 91904, Israel*

[2]*Department of Molecular Chemistry and Materials Science, Weizmann Institute of Science, Rehovot 7610001, Israel*

[*]*Corresponding authors*

*Email: uri.banin@mail.huji.ac.il; dan.oron@weizmann.ac.il*

# Table of contents

**Section S1. Supplementary figures**

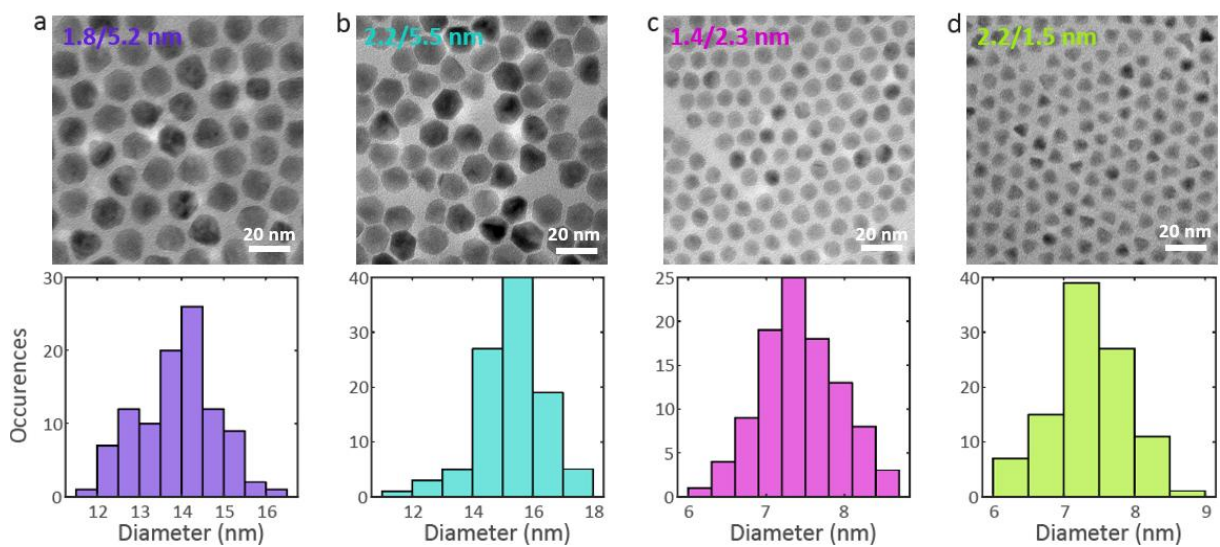

**Figure S1. Size distributions of the studied quantum dots.** The top panels present the transmission electron microscope images of the quantum dots (QDs) that were used in this study. The sizes stated on the top panels showcase the core radii and shell thicknesses. The bottom panels present the size distributions.

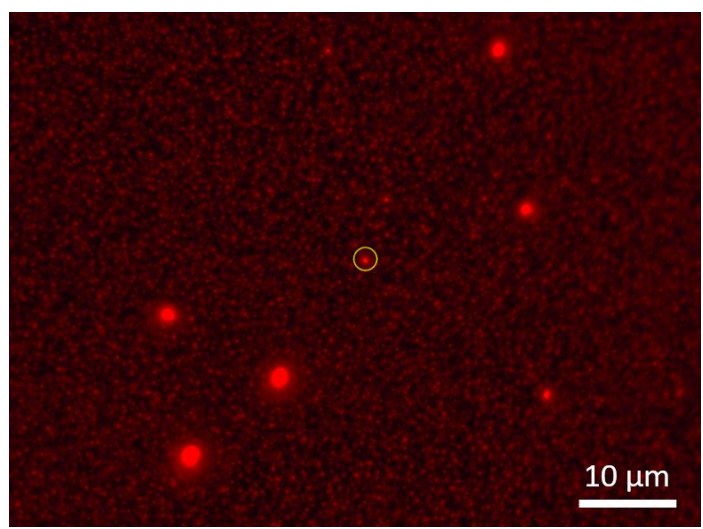

**Figure S2. Widefield photoluminescence image of 1.8/5.2 nm CdSe/CdS QDs.** A widefield photoluminescence image presenting the dispersion of QDs and QD aggregates on the glass substrate. The bright spots are QD aggregates. The yellow circle marks an aggregate that was measured in this study.

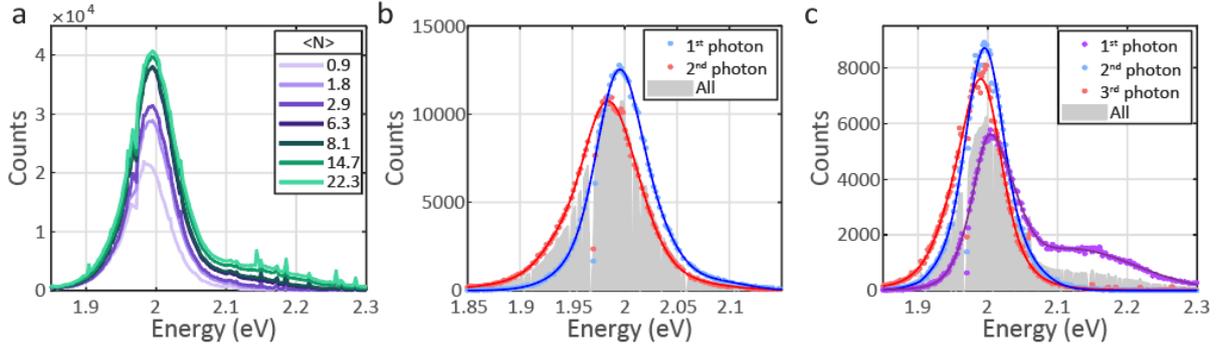

**Figure S3.** Characterization of the aggregate in Figure 1c in the main text. (a) Power dependent spectrum of the aggregate. The ⟨N⟩ values were calculated similarly to the values in Figure 2e in the main text. (b) Spectral characterization of the photon pair events at ⟨N⟩ = 0.9. The blue dots represent the first photons in the pairs and the pink dots represent the second photons in the pairs. They are both fitted to a multi-gaussian model (in blue and red, respectively). The gray area is the normalized spectrum of all photon events. (c) Spectral characterization of the photon triplet events at ⟨N⟩ = 22.3. The first, second, and third photons are in purple, blue, and pink dots, respectively, and are fitted to a multi-gaussian model. The gaps in the spectra are due to excluded noisy pixels. The temporal conditions for the arrival times of the photons in panels b and c are as stated in the Methods section in the main text.

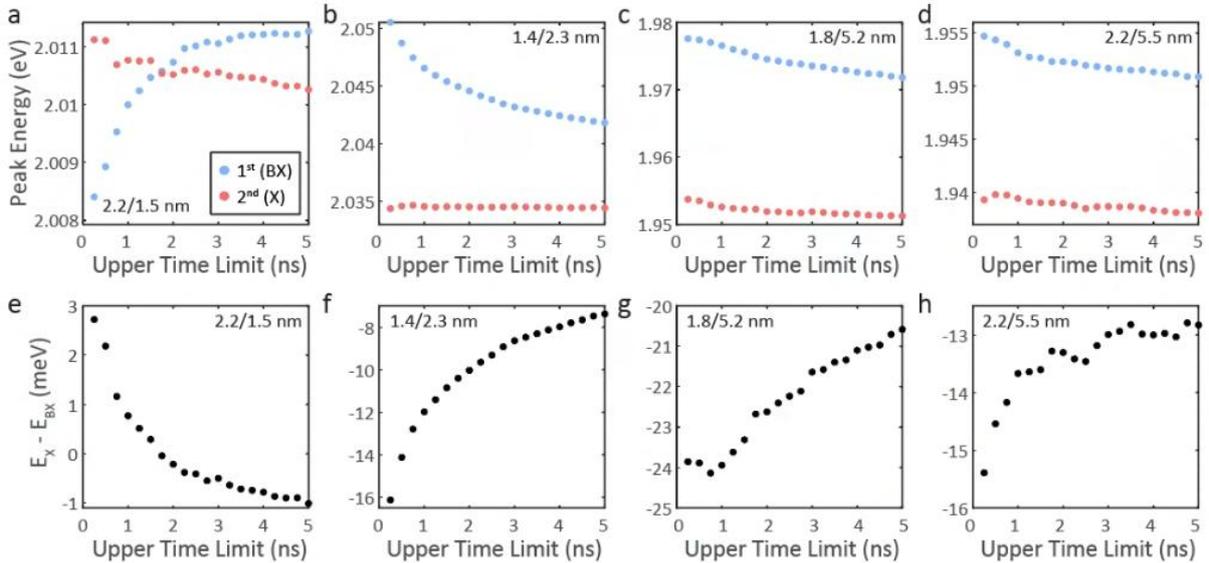

**Figure S4.** (a)-(d) Peak energies of the first (blue) and second (pink) photons within the photon pair events versus the upper limit of the time gate of the first photons (the lower limit is t=0) in the various core/shell sizes that were studied. The second photons are restricted to arrive after the first photons and up to 200 ns after the arrival of the first photons. (e)-(h) are the energy differences between the second and first photons versus the upper time limit for the first

photons in the various studied samples. This analysis showcases that applying a short time gate for the arrival of the first photons in the pairs helps to increase the relative contribution of biexcitons (BXs), apparent by the higher BX energy in short times (or lower energy in the case of the 2.2/1.5 nm QDs, where the BX is red shifted). The time gate presented in the main text is 1 ns. This analysis shows that small changes in the choice of the temporal gate do not change the energies significantly (by few meV at most). The reported BX energy in the main text is extracted from the first data point of this analysis (the energy of the first photons, arriving in the first 0.25 ns after the laser pulse). The error values, presented for the BX binding energies in the main text, reflect the minor influence of this selection, among other contributions.

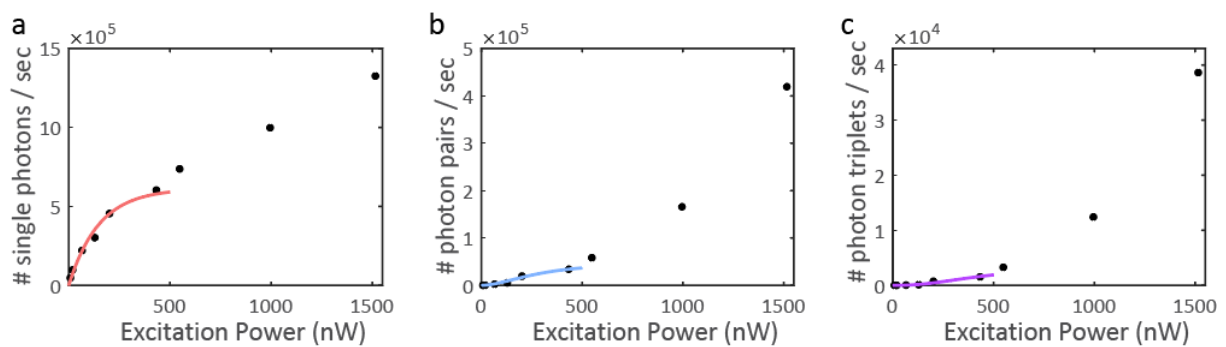

**Figure S5.** The number of single photon events, photon pair events, and photon triplet events per second versus excitation power in panels a, b, and c, respectively (in black dots). The colored lines are fits to the probability to generate at least one, two, or three excitons (in pink, blue, and purple, respectively) according to the Poisson distribution. This is the same analysis as in Figure 2e in the main text, yet it presents the points in the three higher excitation powers that were neglected in the main text. This sharp incline in the emission intensity is due to transition to the quasi-continuous-wave regime, in which the multi-excited QDs emit rapidly during the laser pulse.[1]

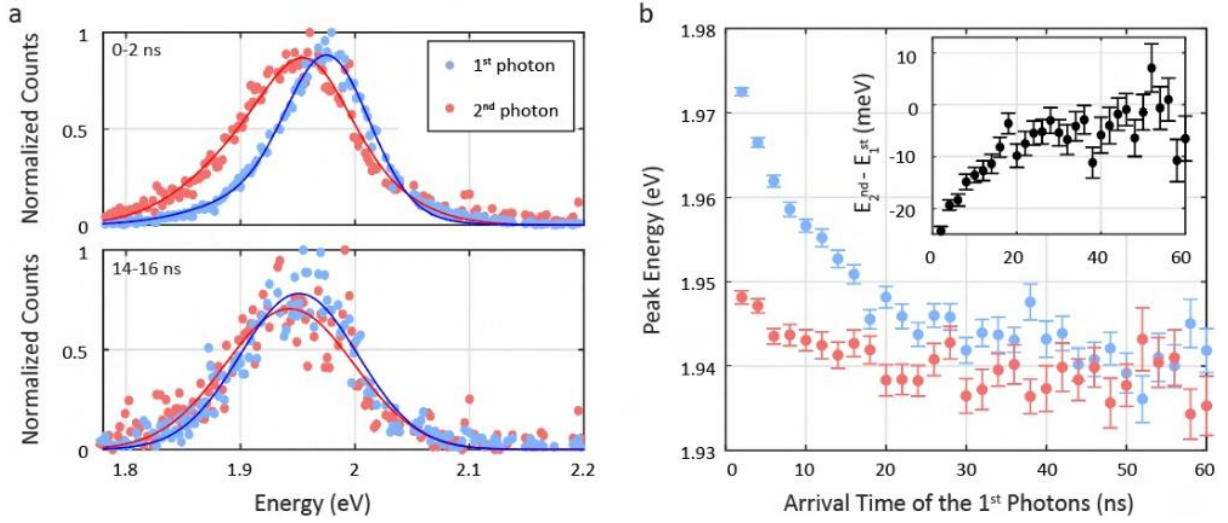

**Figure S6. Biexciton time gated analysis.** This analysis is for the same aggregate of QDs as in Figures 1h-j, 2-3, 5 in the main text, at ⟨N⟩ = 0.1. (a) The spectrum of the first and second photons (in blue and pink, respectively) within the photon pair events. In the top (bottom) panel the arrival time of the first photons is set to 0-2 ns (14-16 ns), and the spectrum of the first photons exhibits a blue shift (no shift) relative to the spectrum of the second photons. The shift in short times represents the interaction within the BX state, whereas after 14 ns the majority of the BXs decayed, and the photon pair events include mostly uncorrelated single excitons (Xs) emitted from different QDs. The spectra are fitted to a multi-gaussian model. (b) Time-dependent peak energies of the spectrum of the first photons (in blue) and of the spectrum of the second photons (in pink), in bins of 2 ns for the arrival time of the first photons in the pairs (the Xs arrive later, between 0.8 and 200 ns after the first photons). The black dots in the inset are the difference between the peak energies of the first and second photons' spectra. The first photons (in blue) exhibit a significant red shift with time, due to the transition from BX-dominated emission to X-dominated emission. Notably, even in longer times, after the BXs decayed, the first photon is somewhat higher in energy than the second photon. This is assigned to the size dispersion in the ensemble, as generally smaller QDs emit faster and in higher energy than larger QDs.[2] Interestingly, the X emission peak in pink slightly red shifts with time as well. This observation is unlikely to stem from the BX emission. For this end, we followed the emission of a single QD in low excitation power (Figure S7).

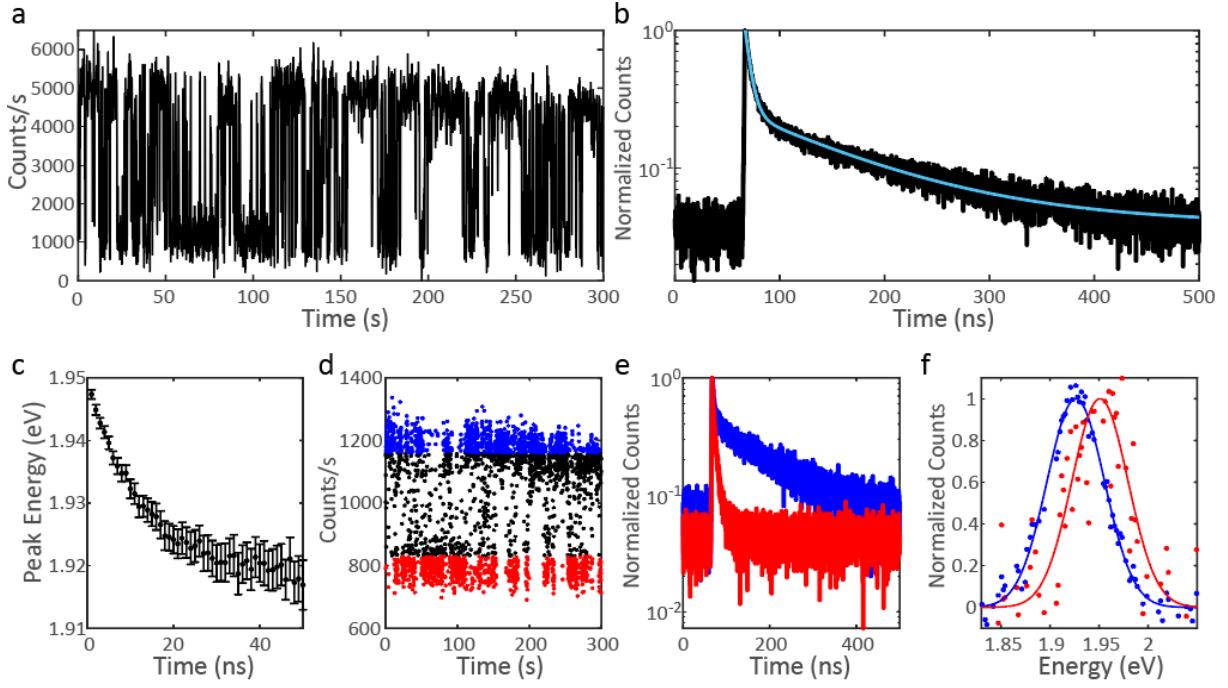

**Figure S7. Characterization of a single QD.** (a) Fluorescence intensity fluctuation time-trace of a single QD at $\langle N \rangle = 0.1$ (excited with an EPL450 pulsed laser by Edinburgh Instruments). The transition between an emissive "on" state and a non-emissive "off" state (i.e. blinking) is indicative of emission of single QDs.[3] (b) Fluorescence decay histogram fitted to a bi-exponential decay (in light blue), showcasing short and long lifetime components (5.3±0.1, 113±1 ns). (c) The peak energy of the emission spectrum versus the arrival time of the photons (in bins of 1 ns), showcasing a red shift in longer times. (d) To understand this behavior in the single QD, we studied the top 30% of the counts (i.e., the "on" state, in blue) and the bottom 30% of the counts (i.e., the "off" state, in red) within the fluorescence intensity fluctuation time-trace. The "on" state is usually associated with neutral X emission, whereas the "off" state is usually associated with charged Xs or even multiexcitons (MXs), that undergo Auger recombination.[4] (e) Fluorescence decay histograms of the "on" state (in blue) and of the "off" state (in red). The decay histogram of the "off" state exhibits only contribution of the short lifetime component, in line with Auger recombination of charged Xs. The decay histogram of the "on" state exhibits a dominant contribution of the long lifetime component, in line with neutral X emission. (f) Normalized emission spectra of the top 30% of the counts (in blue) and the bottom 30% of the counts (in red), fitted to a gaussian distribution. The emission peaks of the "on" and "off" states are 1.926±0.001 and 1.951±0.002, respectively. This indicates that the neutral X emission in a single QD is red shifted by ~25 meV than the charged X emission.

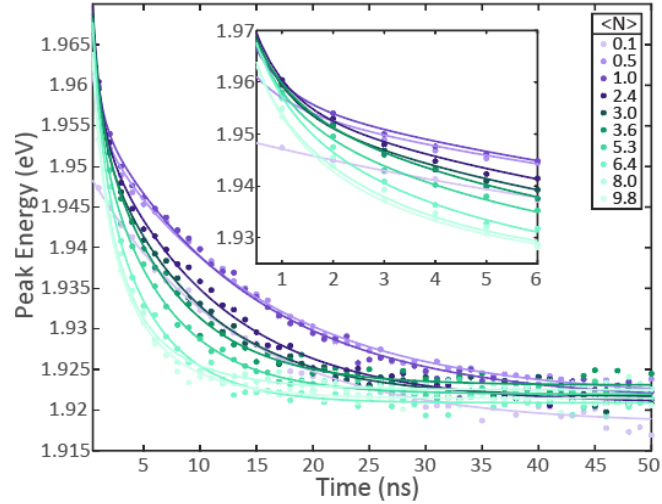

**Figure S8. Power dependent energy shift with time in a single QD.** The peak energy of the emission spectrum versus the arrival time of the photons (in bins of 1 ns) in various excitation powers, in the same single QD as in Figure S7. The $\langle N \rangle$ values are calculated according to the Poisson distribution (similarly to the analysis in Figure 2e in the main text). This power dependent representation of the red shift of the spectrum with time is crucial in order to fully understand the source of the shift. In Figure S7 the "off" state, with fast-emitting photons, is blue shifted relative to the later arriving photons. This could indicate contribution of Auger recombination of either charged Xs or MXs. This figure shows that in the low excitation power of $\langle N \rangle = 0.1$ (with its elaborated results in Figure S7) the peak energy in short times is 1.947 eV. The inset, which zooms in to the first 6 ns of this analysis clearly shows that when increasing the excitation power, a further blue shift is apparent in short times. For example, at $\langle N \rangle = 0.5$ the peak energy in short times is 1.957 eV. In conclusion, the further blue shift in higher excitation powers suggests BX emission (or emission of MXs of higher orders in high powers). However, at $\langle N \rangle = 0.1$ the probability to generate MXs is low, and thus the blue shift in short times is dominated by emission of charged Xs.

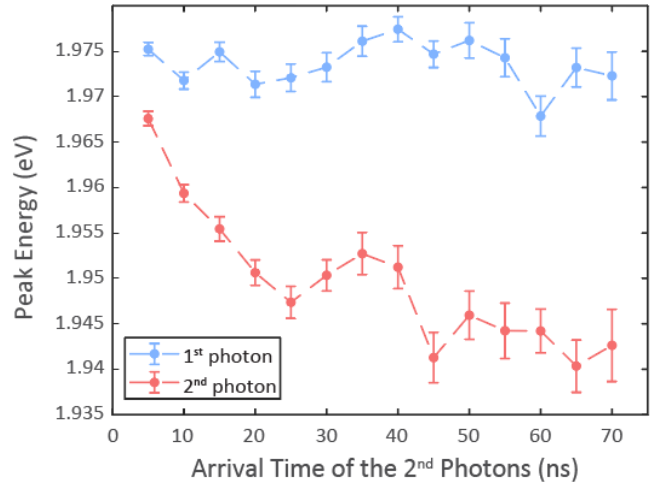

**Figure S9.** The peak energies of the spectrum of the first photons (in blue) and of the spectrum of the second photons (in pink), where the x axis represents the arrival time of the second photons in the pairs (in 5 ns bins). Accordingly, the energy of the first photons does not change significantly, as it includes only photons that arrive in the first 3 ns in all the data points. However, the emission peak of the second photons red shifts with time, due to the decreasing contribution of charged excitons in longer times. Following this analysis, the emission energy of the neutral X was determined as 1.942 eV, as the emission energy of the second photons in long arrival times.

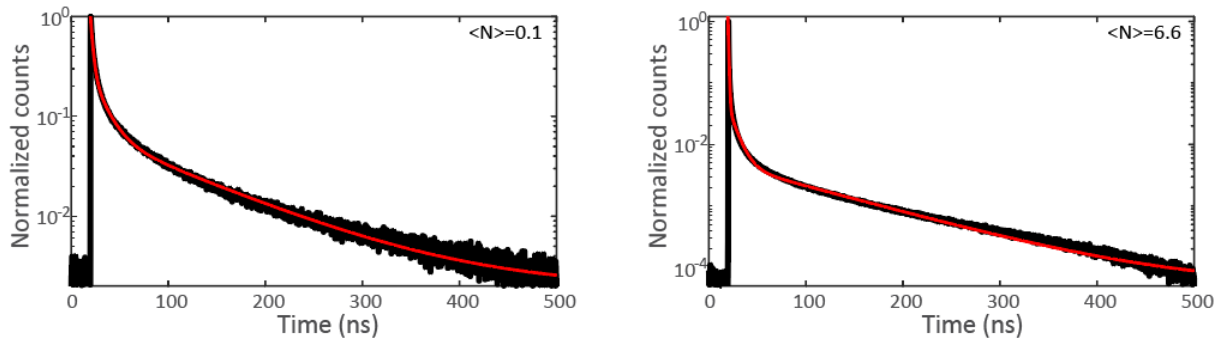

**Figure S10.** Fluorescence decay histograms at excitation powers of $\langle N \rangle = 0.1$ (left panel) and $\langle N \rangle = 6.6$ (right panel) of the same ensemble of QDs as in Figures 1h-j, 2-3, 5 in the main text. Each of the histograms are fitted to tri-exponential decays, where the long component represents the neutral X emission, the short component represents MX emission, and the intermediate component likely represents charged X emission.[5] The fitted lifetimes of the weakly excited aggregate (in the left panel) are 2.4±0.1, 13.8±0.2, 105±1 ns, and the fitted lifetimes of the strongly excited aggregate (in the right panel) are 0.70±0.01, 8±1, 100±45 ns.

The large error value in the long lifetime component is due to the low contribution of the long component to the decay behavior (less than 1%).

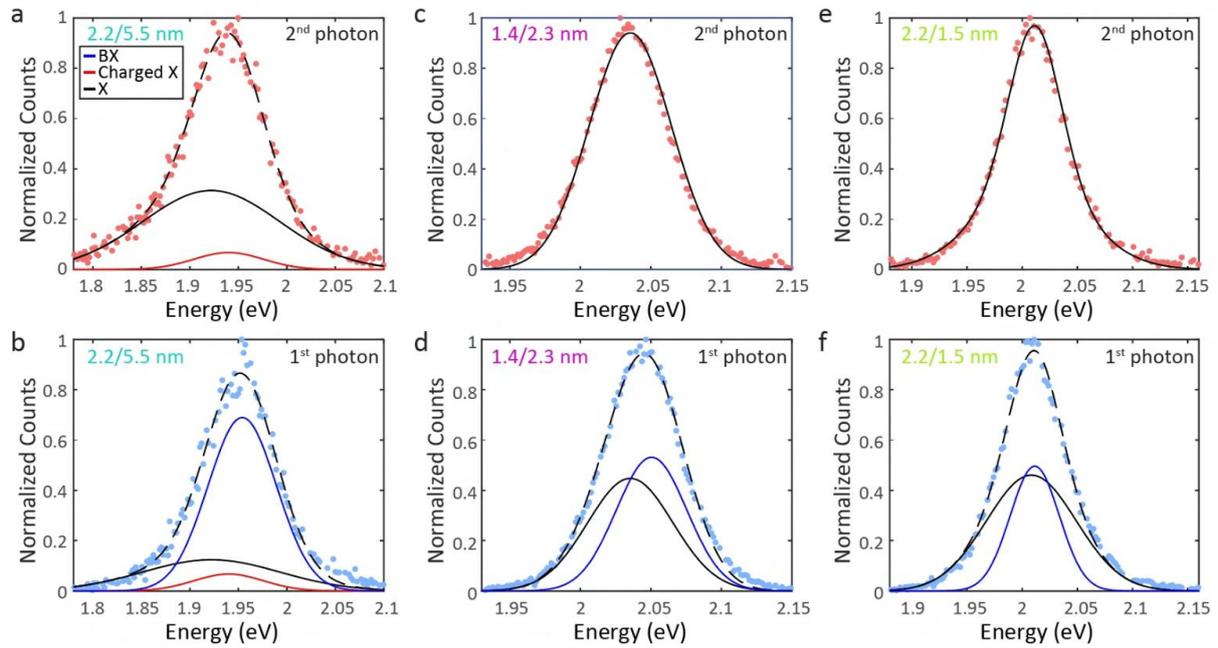

**Figure S11. Size dependent biexciton analysis.** The spectrum of the second and first photons (top panels in pink dots and bottom panel in blue dots, respectively) within the photon pair events for 2.2/5.5 nm QDs (in a, b), 1.4/2.3 nm QDs (in c, d), and 2.2/1.5 nm QDs (in e, f). Only pairs where the first photons arrived in the first 2 ns are included. The second photons in (a) are fitted to a sum of two gaussians (in a dashed black line). Its components represent neutral X emission (black line) and charged X emission (red line). Similarly, the first photons' spectrum in (b) is fitted to a sum of three gaussians (in a dashed black line). The black and red components are the same as in (a), representing neutral and charged X emission, respectively. The blue line represents BX emission. This analysis follows the same procedure for the 1.8/5.2 nm QDs presented in Figure 3 in the main text. The energies of the emitting states are 1.922 (X), 1.94 (charged X), and 1.954 (BX). For the smaller QDs in panels c-f fits of a single gaussian for the second photons and a sum of two gaussians for the first photons are sufficient. The energies of the emitting states in the 1.4/2.3 nm QDs (in c, d) are 2.035 (X), 2.051 (BX). The energies of the emitting states in the 2.2/1.5 nm QDs (in e, f) are 2.011 (X), 2.008 (BX).

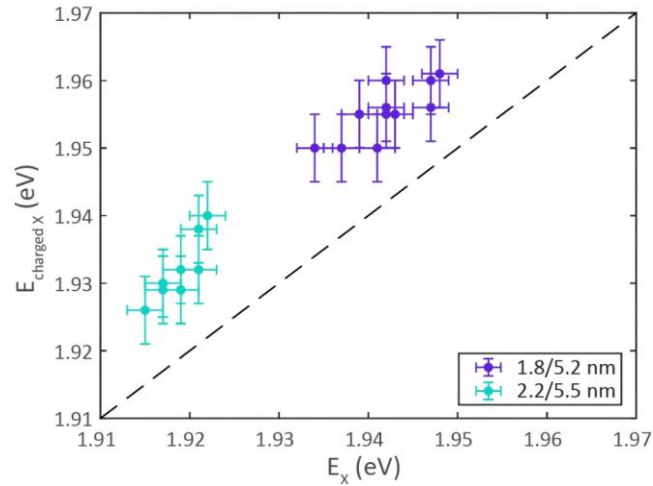

**Figure S12. Charged exciton energies.** Summary of the charged X energies extracted from the BX analysis (as in Figure 3 in the main text and Figure S11a, b) for the 1.8/5.2 and 2.2/5.5 nm QDs at $\langle N \rangle \sim 0.1$. All charged X energies are blue shifted relative to the neutral X energies (and thus appear above the dashed diagonal line, which is used as a guideline to equal X and charged X energies).

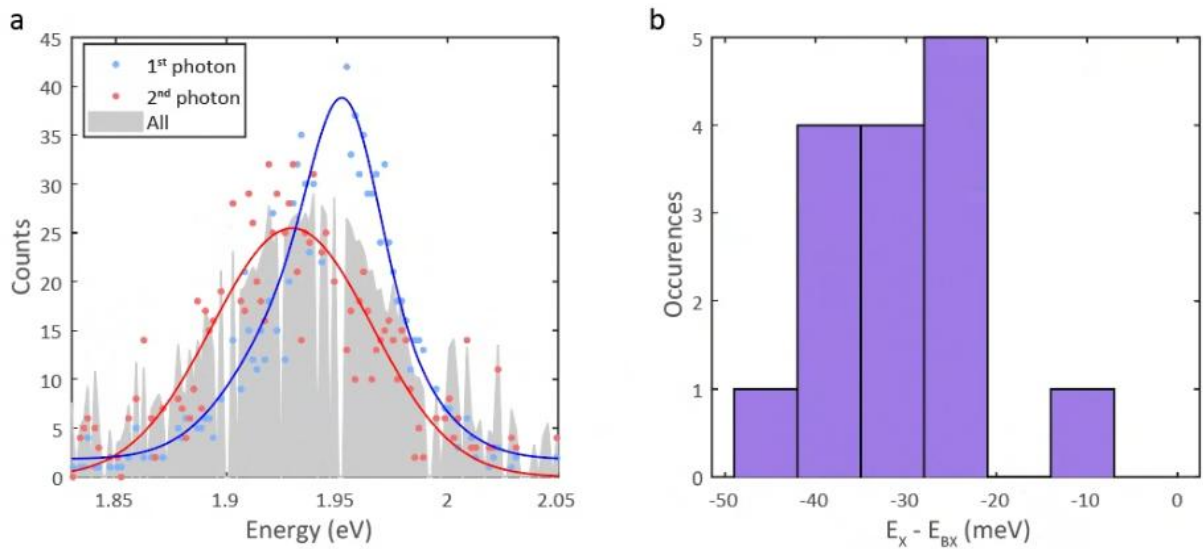

**Figure S13. Biexciton binding energy of single 1.8/5.2 nm QDs.** (a) An example of heralded spectroscopy analysis in a single QD. The blue dots represent the first photons in the pairs and the pink dots represent the second photons in the pairs. They are both fitted to a multi-gaussian model (in blue and red, respectively). The gray area is a normalized spectrum of all photon events. The difference between the X and BX peaks is -22.5 meV. (b) A histogram of the BX binding energy, measured for 15 single QDs. The average BX binding energy is -31±9 meV.

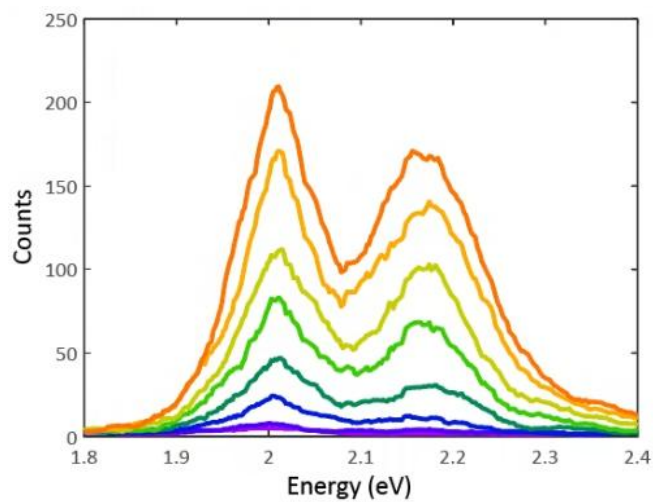

**Figure S14.** The buildup of the spectrum of the first photons within the photon triplet events in time bins of 20 ps, going from purple to orange (followed by the decay of the spectrum in Figure 5c in the main text).

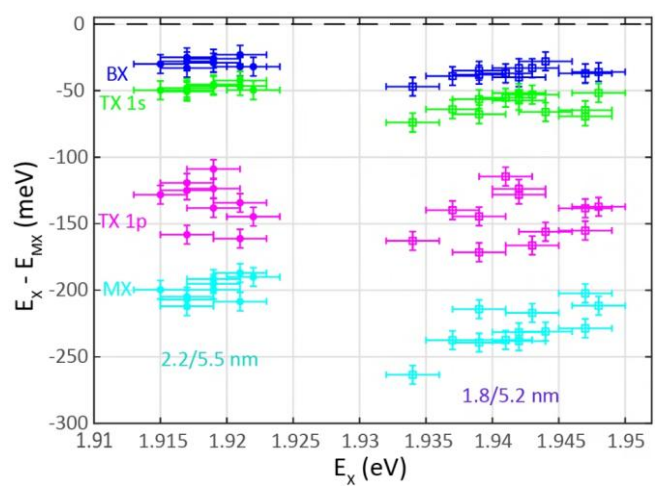

**Figure S15. Size dependent multiexciton energies.** The difference between the peak energy of the X and that of the BX (in blue), the two triexciton (TX) states (in green and magenta), and the high-order MX (in cyan), for all the ensembles from QDs of 2.2/5.5 nm (on the left side) and of 1.8/5.2 nm (on the right side). A negative value indicates a blue shifted state relative to the X energy. The aggregates from the smaller QDs (2.2/1.5, 1.4/2.3 nm) did not exhibit additional emitting states, aside from the BX. This is due to the strong volume dependence of the Auger recombination rate, which significantly lowers the quantum yield of the multi-excited states in the smaller QDs, rendering them hard to detect even in ensemble spectroscopy.[6,7] In the QDs of 2.2/5.5, 1.8/5.2 nm the trend of the BX binding energies persists in the higher order emitting states, such that the energies of the states in the smaller 1.8/5.2 nm

QDs are slightly higher than in the bigger 2.2/5.5 nm QDs, due to the size-dependent quantum confinement effect.[8]

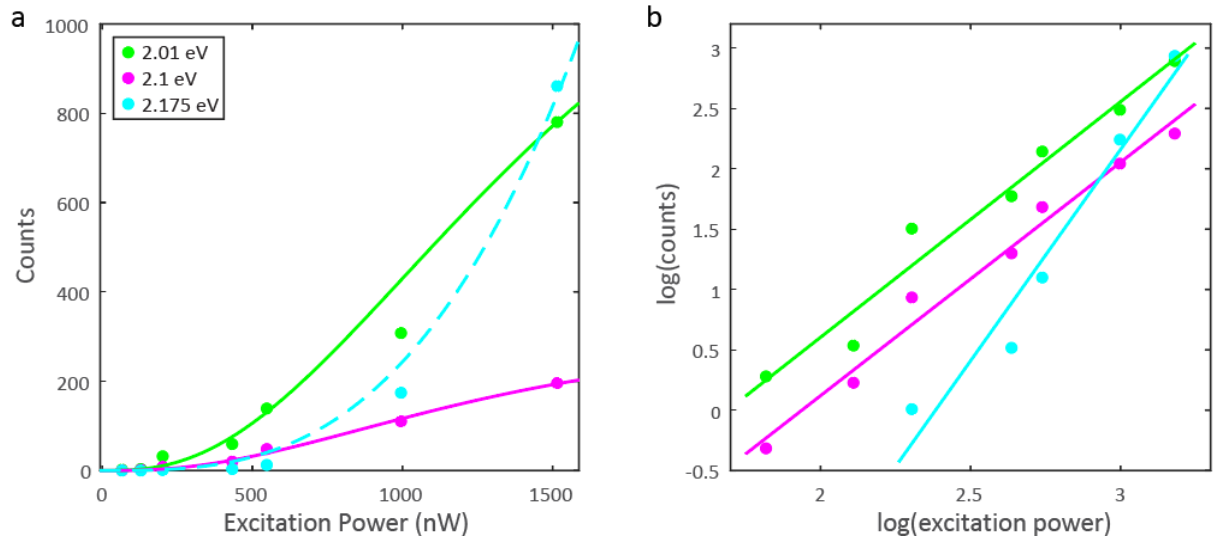

**Figure S16. Power dependent triexciton emission.** (a) The integrated area under each of the three higher energy gaussian components, as presented in Figure 5a in the main text, versus the excitation power. Each curve is fitted to the probability to generate at least three Xs.[9] The curves that describe the power dependence of the 2.01, 2.1 eV (TX) components fit well to the probability to generate at least three Xs, whereas the component at 2.175 eV (high order MX) showcases a different behavior and does not fit well to this function. The different behavior of the 2.175 eV component becomes clearer in a log-log plot of the integrated area under the gaussian fits and the excitation power, as presented in (b). The two lower energy components (2.01 eV in green and 2.1 eV in magenta) exhibit similar slopes (2.0±0.2 and 1.9±0.2, respectively), whereas the 2.175 eV component in cyan exhibits a significantly higher slope (3.5±0.7), suggesting higher order of MX emission.

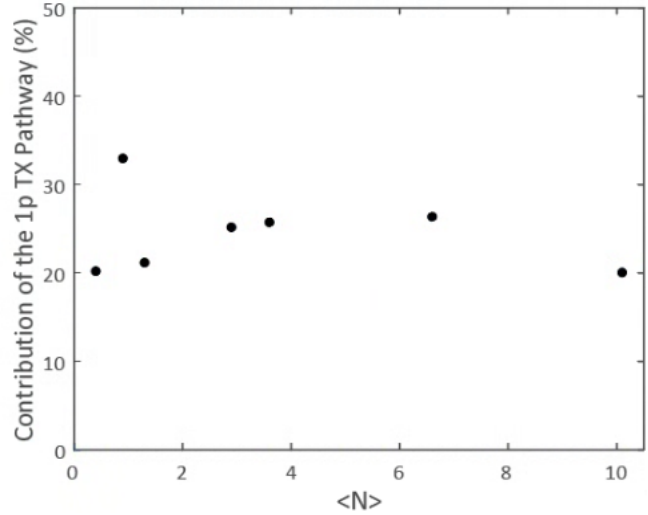

**Figure S17.** The observed relative contribution of the $1p$ TX pathway to the TX emission signal versus the average number of generated Xs per QD per pulse, $\langle N \rangle$. This figure reflects the independence between the contribution of the $1p$ TX emission and the excitation fluence.

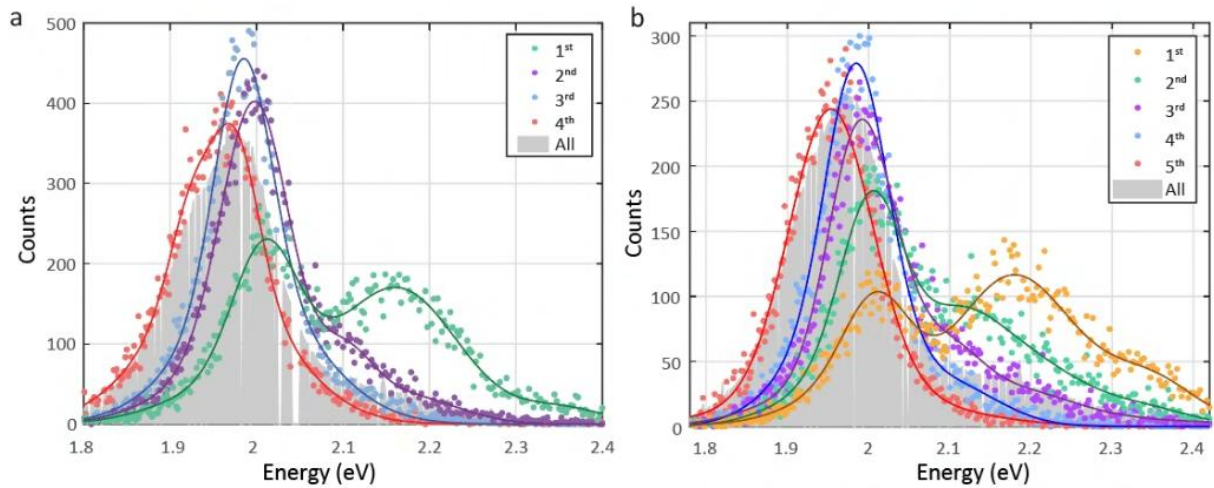

**Figure S18. Heralded spectroscopy analysis for four and five sequential photon events.** This analysis is for the same ensemble of QDs as in Figures 1h-j, 2-3, 5 in the main text, at $\langle N \rangle = 6.6$ (a) and $\langle N \rangle = 10.1$ (b). (a) Spectra of events where four photons are detected after an excitation pulse. The spectrum of the first, second, third, and fourth photons are in green, purple, blue, and pink, respectively. The top gates of arrival time of the first, second, third, and fourth photons are 0.2, 1, 3, and 200 ns. (b) Spectra of events where five photons are detected after an excitation pulse. The top gates of arrival time of the first, second, third, fourth, and fifth photons are 0.3, 0.3, 2, 3, and 200 ns, respectively. The temporal gap between the photon events are 0.8 ns for the 10 nearest neighbor pixels, to avoid cross-talk detection. In farther

pixels, the temporal gap is 0.1 ns (in the 5 photon events, the gap between the first and second photons is even smaller, 0.05 ns). See the Methods section in the main text, which explains the considerations in applying these temporal conditions.

**Section S2. Supplementary text**

Temporal conditions in the time gated TX analysis: Figure 5 in the main text presents the time gated TX analysis. The time gates are as follows: 1. First photon (panel d in the main text) – time gate of 0-2 ns in bins of 20 ps. The time gates for the second and third photons are 3 and 200 ns, respectively. 2. Second photon (panel f in the main text) – time gate of 0.4-20 ns in bins of 20 ps. The time gates for the first and third photons are 0.2 and 200 ns, respectively. 3. Third photon (panel h in the main text) – time gate of 1-20 ns in bins of 50 ps. The time gates for the first and second photons are 0.2 and 3 ns, respectively. The temporal gaps between the photons are 0.8 ns, as explained in the methods section in the main text. The time gated spectra in the upper panels (c, e, g) only showcases part of this data to demonstrate the spectral changes. Therefore, the presented time ranges are shorter and in panels e and g the time bins are larger (as stated in the main text).

**Table S1.** Extracted lifetimes from the time-gated TX analysis in Figure 5 in the main text.

|  | First photon (panel d)[*] | Second photon (panel f) | Third photon (panel h) |
|---|---|---|---|
| Charged exciton | -[**] | 6.1±0.2 ns | 5.4±0.4 ns |
| Biexciton | 1.4±1.3 ns | 1.5±0.1 ns | 3.1±0.4 ns |
| $1s$ triexciton | 323±5 ps | 428±3 ps | - |
| $1p$ triexciton | 227±7 ps | 360±10 ps | - |

[*]The panels refer to Figure 5 in the main text. [**]Missing lifetimes are due to low contribution of these states to the spectrum.

Extracting the lifetimes of the TX pathways: The observed lifetimes for the two TX components were 323±5 and 227±7 ps. These lifetimes reflect the overall rate of the TX transition, and averaging them yields a lifetime of 275 ps. This lifetime can be expressed as:

$$\tau_{observed} = \frac{1}{k_{1s} + k_{1p}}$$

Where $k_i = 1/\tau_i$ are the rate constants of the $i$ TX component ($i = 1p$ or $1s$). Additionally, the $1p$ transition contributed to 25% of the TX emission. Accordingly,

$$\phi_{1p} = \frac{k_{1p}}{k_{1s} + k_{1p}}$$

Where $\phi_{1p}$ is the $1p$ emission fraction.[10] These equations yield the lifetimes of the $1s$ and $1p$ TX transitions: 0.37, 1.1 ns, respectively.